\documentclass[structabstract]{aa}
\usepackage{graphicx}
\usepackage{amsmath}
\usepackage{txfonts}
\begin{document}
   \title{An atmospheric radiation model for Cerro Paranal}

   \subtitle{I. The optical spectral range\thanks{Based on observations made
             with ESO telescopes at Paranal Observatory}}

   \author{S. Noll\inst{1}
   \and    W. Kausch\inst{1}
   \and    M. Barden\inst{1}
   \and    A. M. Jones\inst{1}
   \and    C. Szyszka\inst{1}
   \and    S. Kimeswenger\inst{1}
   \and    J. Vinther\inst{2}
          }

   \institute{Institut f\"ur Astro- und Teilchenphysik, Universit\"at
              Innsbruck, Technikerstr. 25/8, 6020 Innsbruck, Austria\\
              \email{Stefan.Noll@uibk.ac.at}
   \and       European Southern Observatory, Karl-Schwarzschild-Str. 2,
              85748 Garching, Germany
             }

   \date{Received; accepted}


  \abstract
   {}
   {The Earth's atmosphere affects ground-based astronomical observations.
    Scattering, absorption, and radiation processes deteriorate the
    signal-to-noise ratio of the data received. For scheduling astronomical
    observations it is, therefore, important to accurately estimate the
    wavelength-dependent effect of the Earth's atmosphere on the observed
    flux.}
   {In order to increase the accuracy of the exposure time calculator of the
    European Southern Observatory's (ESO) Very Large Telescope (VLT) at Cerro
    Paranal, an atmospheric model was developed as part of the Austrian ESO
    In-Kind contribution. It includes all relevant components, such as
    scattered moonlight, scattered starlight, zodiacal light, atmospheric
    thermal radiation and absorption, and non-thermal airglow emission. This
    paper focuses on atmospheric scattering processes that mostly affect the
    blue ($< 0.55\,$\,$\mu$m) wavelength regime, and airglow emission lines and
    continuum that dominate the red ($> 0.55\,$\,$\mu$m) wavelength regime.
    While the former is mainly investigated by means of radiative transfer
    models, the intensity and variability of the latter is studied with a
    sample of 1186 VLT FORS\,1 spectra.}
   {For a set of parameters such as the object altitude angle, Moon-object
    angular distance, ecliptic latitude, bimonthly period, and solar radio
    flux, our model predicts atmospheric radiation and transmission at a
    requested resolution. A comparison of our model with the FORS\,1 spectra
    and photometric data for the night-sky brightness from the literature,
    suggest a model accuracy of about 20\%. This is a significant improvement
    with respect to existing predictive atmospheric models for astronomical
    exposure time calculators.}
   {}


   \maketitle
%

\newcommand{\radunits}
{$\mathrm{photons\,\,s}^{-1}\,\mathrm{m}^{-2}\,\mu\mathrm{m}^{-1}\,
\mathrm{arcsec}^{-2}$}

\section{Introduction}\label{sec:introduction}

\begin{table*}
\caption[]{Sky model parameters for optical wavelength range}
\label{tab:modelpar}
\centering
\vspace{5pt}
\begin{tabular}{c l l l c c l}
\hline\hline
\noalign{\smallskip}
Parameter\tablefootmark{\,a} & Description &
Unit & Range & Default\tablefootmark{\,b} & Demo run\tablefootmark{\,c} &
Section \\
\noalign{\smallskip}
\hline
\noalign{\smallskip}
$90\degr - z_0$ & altitude of target above horizon &
deg & [0,90] & 90. & 85.1 & \ref{sec:extinction} $-$ \ref{sec:airglow} \\
$\alpha$ & separation of Sun and Moon as seen from Earth &
deg & [0,180] & 0. & 77.9 & \ref{sec:moon} \\
$\rho$ & separation of Moon and target &
deg & [0,180] & 180. & 51.3 & \ref{sec:moon} \\
$90\degr - z_\mathrm{moon}$ & altitude of Moon above horizon &
deg & [-90,90] & -90. & 41.3 & \ref{sec:moon} \\
$d_\mathrm{moon}$ & relative distance to Moon (mean = 1) &
-- & [0.945,1.055] & 1. & 1.\tablefootmark{d} & \ref{sec:moon} \\
$\lambda - \lambda_{\sun}$ & heliocentric ecliptic longitude of target &
deg & [-180,180] & 135. & -124.5 & \ref{sec:zodiac} \\
$\beta$ & ecliptic latitude of target &
deg & [-90,90] & 90. & -31.6 & \ref{sec:zodiac} \\
$S_\mathrm{10.7\,cm}$ & monthly-averaged solar radio flux at $10.7$\,cm &
sfu\tablefootmark{\,e} & $\ge 0$ & 130. & 205.5 &
\ref{sec:data} $-$ \ref{sec:cont} \\
$P_\mathrm{season}$ & bimonthly period (1: Dec/Jan, ..., 6: Oct/Nov; 0: entire
year) &
-- & [0,6] & 0 & 4 & \ref{sec:transmission}, \ref{sec:lines}, \ref{sec:cont} \\
$P_\mathrm{time}$ & period of the night ($x/3$ of night, $x$ = 1,2,3; 0:
entire night) &
-- & [0,3] & 0 & 3 & \ref{sec:lines}, \ref{sec:cont} \\
vac/air & vacuum or air wavelengths &
-- & vac/air & vac & air &
\ref{sec:transmission}, \ref{sec:lines}, \ref{sec:cont} \\
\noalign{\smallskip}
\hline
\end{tabular}
\tablefoot{\\
\tablefoottext{a}{We neglect temperature and emissivity of telescope and
instrument, because these parameters are irrelevant for the optical spectral
range.}\\
\tablefoottext{b}{Used for Table~\ref{tab:refmag}.}\\
\tablefoottext{c}{Used for Figs.~\ref{fig:logcomp}, \ref{fig:compscatextrad},
\ref{fig:varclasses}, and \ref{fig:fors_0135}.}\\
\tablefoottext{d}{Fixed to default value because of its minor importance (see
also Sect.~\ref{sec:moon}).}\\
\tablefoottext{e}{1\,sfu = 0.01\,MJy.}
}
\end{table*}

Ground-based astronomical observations are affected by the Earth's atmosphere.
Light from astronomical objects is scattered and absorbed by air molecules
and aerosols. This extinction effect can cause a significant loss of flux,
depending on the wavelength and weather conditions. The signal of the targeted
object is further deteriorated by background radiation, which is caused by
light from other astronomical radiation sources scattered into the line of
sight and emission originating from the atmosphere itself. Since these
contributions can vary significantly with time, the achievable signal-to-noise
ratio for an astronomical observation strongly depends on the state of the
Earth's atmosphere and the Sun-Earth-Moon system. Therefore, for efficient
time management of any modern observatory, it is critical to provide a reliable
model of the Earth's atmosphere for estimating the exposure time required to
achieve the goals of scientific programmes. Data calibration and reduction also
benefit from a good knowledge of atmospheric effects (see e.g. Davies
\cite{DAV07}).

For this reason, various investigations were performed to characterise the
atmospheric conditions at telescope sites (see Leinert et al. \cite{LEI98} for
a comprehensive overview). Photometric measurements of the night-sky brightness
and its variability were done at e.g. Mauna Kea (Krisciunas \cite{KRI97}), La
Palma (Benn \& Ellison \cite{BEN98}), Cerro Tololo (Walker \cite{WAL87};
Krisciunas et al. \cite{KRI07}), La Silla (Mattila et al. \cite{MAT96}), and
Cerro Paranal (Patat \cite{PAT03}, \cite{PAT08}). For Cerro Paranal, Patat
(\cite{PAT08}) also carried out a detailed spectroscopic analysis, and found
that the night sky showed strong variations of more than one magnitude. Also,
the sky brightness depends on the solar activity cycle (Walker \cite{WAL88};
Patat \cite{PAT08}). It is related to the variations of the upper atmosphere
airglow line and continuum emission, which dominate the near-UV, optical, and
near-IR sky emission under dark-sky conditions (Chamberlain \cite{CHA61}; Roach
\& Gordon \cite{ROA73}; Leinert et al. \cite{LEI98}; Khomich et al.
\cite{KHO08}). When the Moon is above the horizon, scattered moonlight
dominates the blue wavelengths (Krisciunas \& Schaefer \cite{KRI91}). A much
weaker, but always present, component is scattered starlight. The distribution
of integrated starlight (Mattila \cite{MAT80}; Toller \cite{TOL81}; Toller et
al. \cite{TOL87}; Leinert et al. \cite{LEI98}; Melchior et al. \cite{MEL07})
and how it is scattered in the Earth's atmosphere has been studied
(Wolstencroft \& van Breda \cite{WOL67}; Staude \cite{STA75}; Bernstein et al.
\cite{BER02}). A significant component at optical wavelengths is the so-called
zodiacal light, solar radiation scattered by interplanetary dust grains mainly
distributed in the ecliptic plane (Levasseur-Regourd \& Dumont \cite{LEV80};
Mattila et al. \cite{MAT96}; Leinert et al. \cite{LEI98}). For a realistic
description of the zodiacal light intensity distribution for ground-based
observations, scattering calculations are also required (Wolstencroft \& van
Breda \cite{WOL67}; Staude \cite{STA75}; Bernstein et al. \cite{BER02}).
Finally, the wavelength-dependent extinction of radiation from astronomical
objects by Rayleigh scattering off of air molecules, Mie scattering off of
aerosols, and absorption by tropospheric and stratospheric molecules was
studied and characterised at different telescope sites, such as La Silla (T\"ug
\cite{TUG77}, \cite{TUG80}; Rufener \cite{RUF86}; Sterken \& Manfroid
\cite{STE92}; Burki et al. \cite{BUR95}), Cerro Tololo (Stone \& Baldwin
\cite{STO83}; Baldwin \& Stone \cite{BALD84}; Guti\'errez-Moreno et al.
\cite{GUT82}, \cite{GUT86}), and Cerro Paranal (Patat et al. \cite{PAT11}).

\begin{figure}
\centering
\includegraphics[width=8.8cm,,clip=true]{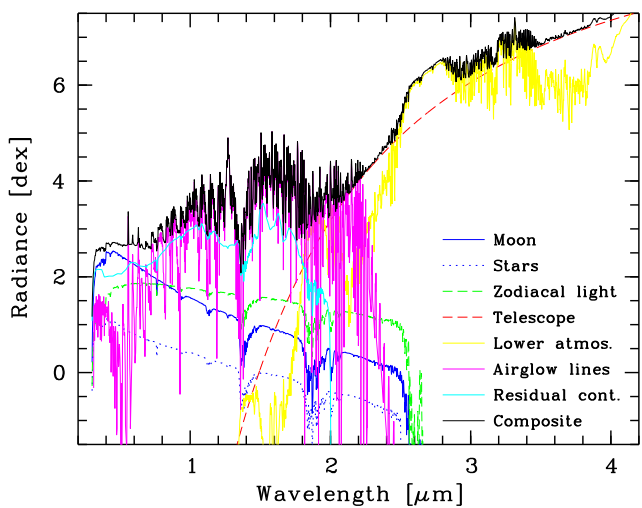}
\caption[]{Components of the sky model in logarithmic radiance units for
wavelengths between 0.3 and 4.2\,$\mu$m. This example, with the Moon above the
horizon, shows scattered moonlight, scattered starlight, zodiacal light,
thermal emission by the telescope and instrument, molecular emission of the
lower atmosphere, airglow emission lines of the upper atmosphere, and
airglow/residual continuum. The scattered light and airglow components were
only computed up to the $K$ band because of their negligible importance at
longer wavelengths. For the model parameters used, see
Table~\ref{tab:modelpar}.}
\label{fig:logcomp}
\end{figure}

Due to the complexity and variability of the night-sky radiation, a good
atmospheric radiation model is crucial for a reliable astronomical exposure
time calculator (ETC). Currently, the European Southern Observatory (ESO) uses
a sky background model for its ETC (Ballester et al. \cite{BALL00}), which
includes the photometric night-sky brightness measurements of Walker
(\cite{WAL87}) for the optical and Cuby et al. (\cite{CUB00}) for the near-IR.
The $U$ to $I$ magnitudes of Walker vary as a function of Moon phase. For
optical spectrographs, a night-sky spectrum is calculated by scaling an
observed, instrument-dependent template spectrum to the Walker filter fluxes.
The spectroscopic near-IR/mid-IR model is based on the line atlas of Hanuschik
(\cite{HAN03}) and the OH airglow calculations of Rousselot et al.
(\cite{ROU00}), thermal telescope emission, an instrument-related constant
continuum, and atmospheric thermal line and continuum emission. The latter is
provided as a set of template spectra computed with the radiative transfer code
Reference Forward Model\footnote{\texttt{http://www.atm.ox.ac.uk/RFM/}} for
different airmasses and water vapour column densities. Since the current ETC is
limited in reproducing the variable intensity of the night sky, we have
developed an advanced model for atmospheric radiation and transmission, which
includes scattered moonlight, scattered starlight, zodiacal light, thermal
emission from the telescope, molecular emission and absorption in the lower
atmosphere, and airglow line and continuum emission, including their
variability with time. This ``sky model'' has been derived for Cerro Paranal in
Chile (2635\,m, 24$\degr$~38$\arcmin$~S, 70$\degr$~24$\arcmin$~W). It is also
expected to work for the nearby Cerro Armazones (3064\,m,
24$\degr$~36$\arcmin$~S, 70$\degr$~12$\arcmin$~W), the future site of the
European Extremely Large Telescope (E-ELT), without major adjustments. An
example of a spectrum computed by our sky model is given in
Fig.~\ref{fig:logcomp}. The crucial input parameters are listed in
Table~\ref{tab:modelpar}. The model will be incorporated into the ESO
ETC\footnote{\texttt{http://www.eso.org/observing/etc/}} and will be made
publicly available to the community via the ESO web site.

This paper focuses on the discussion of model components relevant for the
wavelength range from 0.3 to 0.92\,$\mu$m, which is designated as ``optical''
in the following. We will discuss scattering and absorption in the atmosphere
(Sect.~\ref{sec:extinction}), contributions from extraterrestrial radiation
sources (Sect.~\ref{sec:scatcomp}), and the intensity and variability of
airglow emission lines and continuum (Sect.~\ref{sec:airglow}). The quality of
our optical sky model will be evaluated in Sect.~\ref{sec:discussion}. The
near-IR and mid-IR regimes will be treated in a subsequent paper.

If intensity units are not explicitly given in this paper, \radunits{} are
taken as the standard. Magnitudes are always given in the Vega system.

\section{Atmospheric extinction}\label{sec:extinction}

Light from astronomical objects is scattered and absorbed in the Earth's
atmosphere. For point sources, both effects result in a loss of radiation,
which is usually described by a wavelength-dependent extinction curve or an
atmospheric transmission curve. The components of such a curve for Cerro
Paranal are discussed in Sect.~\ref{sec:transmission}. For extended sources,
light is not only scattered out of the line of sight, it is also scattered into
it. This effect causes an effective extinction curve to differ from that of a
point source and depends on the spatial distribution of the extended emission.
To quantify the change of the extinction curve, we performed three-dimensional
(3D) scattering calculations, which are described in
Sect.~\ref{sec:scattering}.

\subsection{The transmission curve}\label{sec:transmission}

The atmospheric transmission depends on scattering and absorption. Light can be
scattered in dry atmosphere by air molecules such as N$_2$ and O$_2$ or by
aerosols like silicate dust, sea salt, soot, or droplets of sulphuric acid. The
former effect is known as Rayleigh scattering, and is characterised by a strong
wavelength dependence proportional to $\lambda^{-4}$ and a relatively isotropic
scattering phase function (a factor of 2 variation for unpolarised radiation).
The latter can be described by Mie scattering if spherical particles are
assumed. Aerosol scattering is characterised by a relatively weak wavelength
dependence ($\lambda^{-1}$ to $\lambda^{-2}$) and pronounced forward scattering,
for which the maximum intensity can easily be two orders of magnitude higher
than at large scattering angles. Atmospheric absorption in the optical is
mainly caused by bands from three molecules: molecular oxygen (O$_2$), water
vapour (H$_2$O), and ozone (O$_3$). The absorption of O$_2$ is a function of
atmospheric density. While water absorption is most efficient close to the
ground, where the absolute humidity is the largest, ozone mainly absorbs at
stratospheric altitudes of about 20\,km. The combination of scattering and
absorption results in an extinction curve, which is often given in
mag~airmass$^{-1}$, or a transmission curve providing values from 0 (totally
opaque) to 1 (fully transparent). The transmission $t(\lambda)$ can be linked
to the zenithal optical depth $\tau_0(\lambda)$ and zenithal extinction
coefficient $k(\lambda)$ by
\begin{equation}\label{eq:extinction}
t(\lambda) = \mathrm{e}^{-\tau_0(\lambda)\,X} = 10^{\,-\,0.4\,k(\lambda)\,X}.
\end{equation}
The airmass $X$ can be calculated by the formula of Rozenberg
(\cite{ROZ66}):
\begin{equation}\label{eq:airmass}
X = \left(\cos(z) + 0.025 \,\mathrm{e}^{-11 \cos(z)}\right)^{-1},
\end{equation}
where $z$ is the zenith distance and $X$ converges to 40 at the horizon.

\begin{figure}
\centering
\includegraphics[width=8.8cm,clip=true]{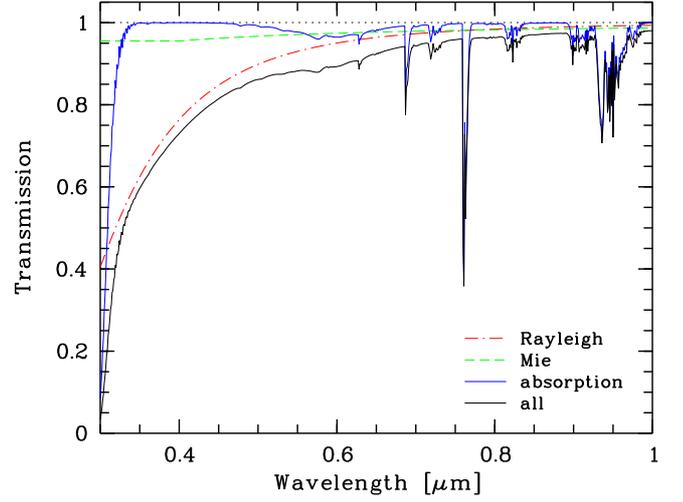}
\caption[]{Annual mean zenith transmission curve for Cerro Paranal (black solid
line). The Earth's atmosphere extinguishes the flux from point sources by
Rayleigh scattering by air molecules (red dash-dotted line), Mie scattering by
aerosols (green dashed line), and molecular absorption (blue solid line). For
the plotted wavelength range, the latter is caused by molecular oxygen (A~band
at 0.762\,$\mu$m, B~band at 0.688\,$\mu$m, and $\gamma$~band at
$0.628$\,$\mu$m), water vapour (prominent bands at 0.72, 0.82, and
0.94\,$\mu$m), and ozone (Huggins bands in the near-UV and broad Chappuis bands
at about 0.6\,$\mu$m).}
\label{fig:transcurve}
\end{figure}

\begin{figure}
\centering
\includegraphics[width=8.8cm,clip=true]{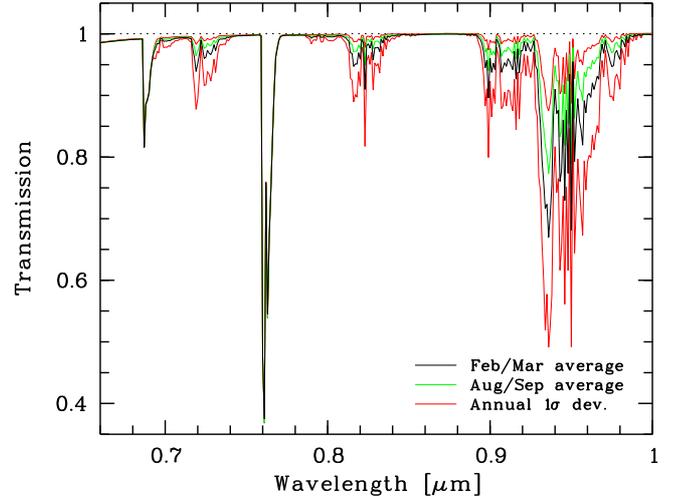}
\caption[]{Variation of molecular absorption for Cerro Paranal. The extreme
bimonthly mean transmission curves and 1\,$\sigma$ deviations of the annual
mean curve (red outer curves) are shown. The highest mean transmission (lowest
water vapour content) is found for August/September (green curve), and the
lowest arises in February/March (black curve). In contrast to H$_2$O, the
variations of the O$_2$ bands are very small. For the identity of the bands,
see Fig.~\ref{fig:transcurve}.}
\label{fig:varwaterabs}
\end{figure}

For Cerro Paranal, Fig.~\ref{fig:transcurve} shows the annual mean transmission
curve at zenith and its components. The extinction at blue wavelengths is
dominated by Rayleigh scattering. This component is very stable and can be well
reproduced by the parametrisation
\begin{eqnarray}\label{eq:rayleigh}
\tau_\mathrm{R}(\lambda) = \frac{p}{1013.25}
\left(0.00864 + 6.5 \times 10^{-6} H\right)
\nonumber\\
\times \ \lambda^{\,-(3.916 \,+ \,0.074 \,\lambda \,+ \,0.050 \,/ \,\lambda)}
\end{eqnarray}
with wavelength $\lambda$ in $\mu$m (see Liou \cite{LIO02}). For the pressure
$p$ and the height $H$, we take 744\,hPa and 2.64\,km respectively. The
pressure corresponds to the annual mean for Cerro Paranal
($743.5 \pm 1.5$\,hPa), as derived from the meteorological station of the VLT
Astronomical Site Monitor.

At red wavelengths, aerosol scattering becomes as important as Rayleigh
scattering. However, the total amount of extinction by scattering is small
in this wavelength regime. For Cerro Paranal, Patat et al. (\cite{PAT11})
provide an approximation for the aerosol extinction derived from 600 VLT
FORS\,1 spectra observed over six months. The aerosol extinction coefficient is
parametrised by
\begin{equation}\label{eq:mie}
k_\mathrm{aer}(\lambda) \approx k_0 \,\lambda^{\,\alpha},
\end{equation}
where $k_0 = 0.013 \pm 0.002$\,mag~airmass$^{-1}$ and $\alpha = -1.38 \pm 0.06$,
with the wavelength $\lambda$ in $\mu$m. Due to an increased discrepancy
between the fit and the observed data in the near-UV (see Patat et al.
\cite{PAT11}), we use Eq.~\ref{eq:mie} only for wavelengths longer than
0.4\,$\mu$m. For shorter wavelengths, we use a constant value of
$k_\mathrm{aer} = 0.050$\,mag~airmass$^{-1}$, which corresponds to the fit value
at 0.4\,$\mu$m. The density, distribution, and composition of aerosols is much
more variable than what is observed for the air molecules, which determine
the Rayleigh scattering. Patat et al. (\cite{PAT11}) find that $k_\mathrm{aer}$
varies by about 20\% at 0.4\,$\mu$m. However, these variations are of minor
importance for the total transmission of the Earth's atmosphere, since the
aerosol extinction coefficients are small.

Figure~\ref{fig:transcurve} exhibits several prominent absorption bands (see
also Patat et al. \cite{PAT11}). At wavelengths below 0.34\,$\mu$m, there is a
conspicuous fall-off of the transmission curve caused by the Huggins bands of
ozone. The stratospheric ozone layer is also responsable for the Chappuis
absorption bands between 0.5 and 0.7\,$\mu$m. The relatively narrow, but
strong, bands at 0.688\,$\mu$m and 0.762\,$\mu$m can be identified as the
Fraunhofer B and A bands of molecular oxygen. Finally, the complex bands at
0.72, 0.82, and 0.94\,$\mu$m are produced by water vapour.

The molecular absorption bands have been calculated using the Line By Line
Radiative Transfer Model (LBLRTM), an atmospheric radiative transfer code
provided by the Atmospheric and Environmental Research Inc. (see Clough et al.
\cite{CLO05}). This widely used code in the atmospheric sciences computes
transmission and radiance spectra based on the molecular line database
HITRAN (see Rothman et al. \cite{ROT09}) and atmospheric vertical profiles of
pressure, temperature, and abundances of relevant molecules.

For Cerro Paranal, we use merged atmospheric profiles from three data sources
to reproduce the climate and weather conditions in an optimal way. First, the
equatorial daytime standard profile from the MIPAS instrument of the ENVISAT
satellite (prepared by J. Remedios 2001; see Seifahrt et al. \cite{SEI10}) is
taken. It provides abundances for 30 molecular species up to an altitude of
120\,km. Following Patat et al. (\cite{PAT11}), the ozone profile is corrected
by a factor of 1.08 to achieve a column density of 258 Dobson units, which
represents the mean value for Cerro Paranal. Second, we use profiles from the
Global Data Assimilation
System\footnote{\texttt{http://ready.arl.noaa.gov/gdas1.php}} (GDAS),
maintained by the Air Resources Laboratory of the National Oceanic and
Atmospheric Administration (cf. Seifahrt et al. \cite{SEI10}). The GDAS
profiles for pressure, temperature, and relative humidity are provided on a
3\,h basis for a $1\degr \times 1\degr$ global grid. These models are adapted
to data from weather stations all over the world and are suitable for
weather-dependent temperature and water vapour profiles up to altitudes of
26\,km. Third, data from the meteorological station at Cerro Paranal are used
to scale the pressure, temperature, and water vapour profiles at the altitude
of the mountain. For higher altitudes, the scaling factor is reduced and
approaches 1 at 5\,km.

For our sky model, we analysed the resulting data set and constructed mean
profiles with their $1\,\sigma$ deviations for different periods. We divide the
year into six two-month periods, starting with December/January (cf.
Table~\ref{tab:modelpar} and Sect.~\ref{sec:airglow}). The two most extreme
mean spectra are shown in Fig.~\ref{fig:varwaterabs}. The seasonal variability
of the H$_2$O bands is clearly visible. For the driest period
(August/September), the mean absorption is only half the amount of the most
humid period (February/March). The total intra-annual variability indicates
line depth variations of an order of magnitude, i.e. large statistical
uncertainties. For this reason, the significant seasonal dependence is included
in the sky model and the less pronounced average, nocturnal variations have
been neglected. Apart from the two-month period, only the airmass (see
Eq.~\ref{eq:airmass}) is used as input parameter for the computation of
transmission curves. Since radiative transfer calculations with LBLRTM are
time-consuming, the sky model is run with a pre-calculated library of
transmission spectra, consisting of spectra for the different bimonthly
periods and a regular grid of five airmasses between 1 and 3. This is
sufficient for a reliable interpolation of the airmass-dependent change of
spectral features.

A more detailed discussion on atmospheric profiles, radiative transfer codes,
and the properties of transmission and radiance spectra, especially in the
near-IR and mid-IR, will be given in a subsequent paper.

\subsection{Atmospheric scattering}\label{sec:scattering}

\begin{figure}
\centering
\includegraphics[width=8.8cm,clip=true]{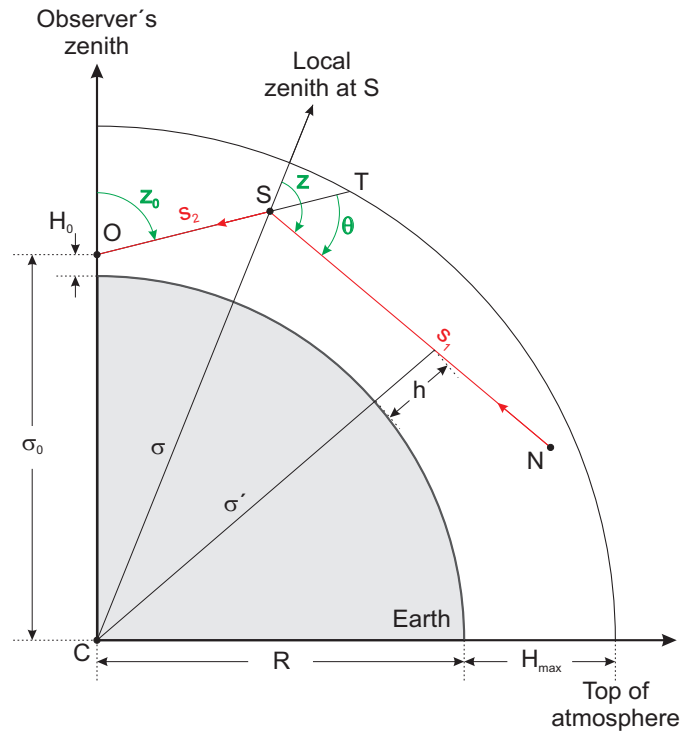}
\caption[]{Geometry of the scattering in the Earth's atmosphere (cf.
Wolstencroft \& van Breda \cite{WOL67}). Point $N$ is at the top of atmosphere
and not in the same plane as the other points. The azimuth of $N$ as seen from
$S$ is $A$ (not shown). $S$ and $T$ are at an azimuth $A_0$ (also not shown)
for an observer at $O$.}
\label{fig:scatsketch}
\end{figure}

The estimate of scattered light from extended sources, such as integrated
starlight, zodiacal light, or airglow in the Earth's atmosphere, requires
radiative transfer calculations. Since the optical depth for scattering is
relatively small in most of the optical wavelength range (see
Fig.~\ref{fig:transcurve}), single-scattering calculations provide a sufficient
approximation. In this case, the computations can be performed in 3D with a
relatively compact code (see Wolstencroft \& van Breda \cite{WOL67}; Staude
\cite{STA75}; Bernstein et al. \cite{BER02}).

To obtain the integrated scattered light towards the azimuth $A_0$ and zenith
distance $z_0$, we consider scattering path elements $S$ of density
$n(\sigma)$, with $\sigma$ being the radius vector from the centre of Earth $C$
to $S$, from the top of atmosphere $T$ to the observer $O$ at height $H_0$
above the surface (see Fig.~\ref{fig:scatsketch}). The distance of $O$ to $C$
is $\sigma_0 = H_0 + R$, where $R$ is the radius of Earth ($= 6371$\,km for the
mean radius). For each path element $S$ at distance $s_2$ from $O$, the
contributions of radiation from all directions ($A$, $z$) to the intensity at
$S$ are considered. The integration over the solid angle depends on the spatial
intensity distribution $I_0(A, z)$ of the extended radiation source and the
path of the light from the entry point in the atmosphere $N$ to $S$ of length
$s_1$. The latter includes the scattering of light out of the path (and
possible absorption), which depends on the effective column density of the
scattering/absorbing particles
\begin{equation}\label{eq:effcoldens}
B_1(z, \,\sigma) = \int_{0}^{s_1(z, \,\sigma)} n(\sigma') \,\mathrm{d}\sigma'.
\end{equation}
It also depends on the wavelength-dependent extinction cross section
$C_\mathrm{ext}(\lambda)$, which can be derived from the optical depth at zenith
$\tau_0$ (see Eq.~\ref{eq:extinction}) by means of Eq.~\ref{eq:effcoldens} and
\begin{equation}\label{eq:crosssection}
C_\mathrm{ext}(\lambda) = \frac{\tau_0(\lambda)}{B_0(0, \,\sigma_0)}.
\end{equation}
After the scattering at $S$, the intensity is further reduced by the effective
column density $B_2(z_0, \sigma)$ along the path $s_2$. Thus, the calculations
can be summarised by
\begin{eqnarray}\label{eq:scatintegral}
I_\mathrm{scat}(A_0, z_0) = \frac{C_\mathrm{scat}(\lambda)}{4\pi}
\int_0^{s_2(z_0, \,\sigma_0)} \int_0^{z_\mathrm{max}(\sigma)} \int_0^{2\pi} n(\sigma)
\,P(\theta) \nonumber\\
\times \ I_0(A, z)
\,\mathrm{e}^{-C_\mathrm{ext}(\lambda)\left[B_1(z, \,\sigma) + B_2(z_0, \,\sigma)\right]}
\,\mathrm{d}A \,\sin z \,\mathrm{d}z \,\mathrm{d}s.
\end{eqnarray}
$C_\mathrm{scat}$ is the wavelength-dependent scattering cross section, which
will deviate from $C_\mathrm{ext}$ if absorption occurs. The maximum
zenith distance $z_\mathrm{max}$ is higher than $90\degr$ and depends on the
height of $S$ above the ground. The scattering phase function $P$ depends on
the scattering angle $\theta$, the angle between the paths $s_1$ and $s_2$.

Rayleigh scattering (see Sect.~\ref{sec:transmission}) is characterised by the
phase function
\begin{equation}\label{eq:ptheta}
P(\theta) = \frac{3}{4} \left(1 + \cos^2(\theta)\right).
\end{equation}
Similar to Bernstein et al. (\cite{BER02}), we neglect the effect of
polarisation on the scattering phase function. Even for zodiacal light, where
some polarisation is expected (e.g. Staude \cite{STA75}), the polarisation does
not appear to significantly affect the integrated scattered light. Results of
Wolstencroft \& van Breda (\cite{WOL67}) suggest deviations of only a few per
cent. For the vertical distribution of the scattering molecules, we use the
standard barometric formula
\begin{equation}\label{eq:baroform}
n(h) = n_0 \,\mathrm{e}^{-h/h_0}.
\end{equation}
Here, $h = \sigma' - R$, the sea level density
$n_0 = 2.67 \times 10^{19}$\,cm$^{-3}$, and the scale height $h_0 = 7.99$\,km
above the Earth's surface (cf. Bernstein et al. \cite{BER02}). For the
troposphere and the lower stratosphere, where most of the scattering occurs,
this is a good approximation. Cerro Paranal is at an altitude of
$H_0 = 2.64$\,km. For the thickness of the atmosphere, we take
$H_\mathrm{max} = 200$\,km. With these values, $C_\mathrm{ext}$ is
$1.75 \times 10^{-26}$\,cm$^2$ for $\tau_0 = 0.27$, which corresponds to a
wavelength of about 0.4\,$\mu$m. For Rayleigh scattering,
$C_\mathrm{scat} = C_\mathrm{ext}$, i.e. no absorption is involved. In the
near-UV, where the zenithal optical depth is greater than 0.3, the single
scattering approximation becomes questionable. For this reason, we apply a
multiple scattering correction derived from radiative transfer calculations in
a plane-parallel atmosphere (see Dave \cite{DAV64}; Staude \cite{STA75}). We
multiply $I_\mathrm{scat}$ by the factor
\begin{equation}\label{eq:multiscat}
F_\mathrm{MS} = 1 + 2.2 \,\tau_0.
\end{equation}
The uncertainty is in the order of 5\%. The factor $F_\mathrm{MS}$ does not
include reflection from the ground. Mattila (\cite{MAT03}) reports a ground
reflectance of about 8\% in the region near the Las Campanas Observatory in
autumn. Since this telescope site is also located in the Atacama desert, we
assume this value with an uncertainty of a few per cent. Using the tables of
Ashburn (\cite{ASH54}), an 8\% reflectance translates into an additional
$I_\mathrm{scat}$ correction factor of about 1.07.

The Mie scattering of aerosols is characterised by a phase function with a
strong peak in forward direction. For this study, we take the measured
$P(\theta)$ of Green et al. (\cite{GRE71}), which covers the peak up to an
angle of $20\degr$. The phase function for larger scattering angles is
extrapolated by a linear fit in the $\log P - \log \theta$ diagram. This simple
approach neglects the increase of the phase function for scattering angles
close to $180\degr$ (see e.g. Liou \cite{LIO02}). However, since $P$ already
decreases by a factor of 30 from $0\degr$ to $20\degr$, the details of the
phase function at larger angles are not crucial for the scattering of extended
emission. The total scattered intensity is completely dominated by the
contribution from angles close to the forward direction. For the height
distribution of aerosols, we also take Eq.~\ref{eq:baroform}, but assume
$n_0 = 1.11 \times 10^{4}$\,cm$^{-3}$ and $h_0 = 1.2$\,km. This is the
tropospherical distribution of Elterman (\cite{ELT66}). Following Staude
(\cite{STA75}), we neglect the stratospheric aerosol component, which contains
only about 1\% of the particles. Using Eq.~\ref{eq:crosssection} we obtain
$C_\mathrm{ext} = 3.39 \times 10^{-10}$\,cm$^2$ for $\tau_0 = 0.05$, which
characterises wavelengths of about 0.4\,$\mu$m. Dust, in particular soot,
also absorbs radiation. For this reason, $C_\mathrm{scat}$ is lower than
$C_\mathrm{ext}$. The ratio of these is called the single scattering albedo
$\tilde{\omega}$. Following the recommendation of Mattila (\cite{MAT03}),
$\tilde{\omega} = 0.94$ is used for the model, i.e. strong absorbers like
soot do not significantly contribute to the aerosol population. We neglect any
corrections for multiple scattering and ground reflection for Mie scattering,
since the optical depths are low at all relevant wavelengths (see
Fig.~\ref{fig:transcurve}) and forward scattering dominates.

\begin{figure}
\centering
\includegraphics[width=8.8cm,clip=true]{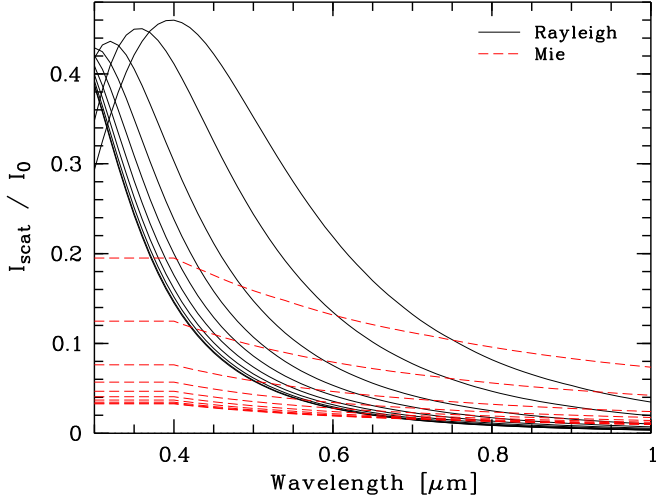}
\caption[]{Intensity of scattered radiation for a uniform source of unit
brightness using the Paranal extinction curve (see
Sect.~\ref{sec:transmission}). Contributions of scattered light into the line
of sight are shown for Rayleigh (solid lines) and Mie scattering (dashed lines)
and zenith distances from 0$\degr$ to 80$\degr$ in 10$\degr$ steps plus an
extreme value of 85$\degr$. Except for Rayleigh scattering at very short
wavelengths, the scattering intensities increase with zenith distance.}
\label{fig:fracscatunit}
\end{figure}

To test our model, we computed scattering intensities for a uniform radiation
source of unit brightness. For the surface level and $\tau_0 = 0.052$, we
obtain scattering fractions of $0.024$ for the zenith and $0.360$ for the
horizon ($z_0 = 90\degr$) with the multiple scattering corrections neglected.
For Rayleigh scattering, these results can be compared to those of Staude
(\cite{STA75}). Our values are only about 5\% higher. This is a good agreement
compared to differences between Staude (\cite{STA75}) and older calculations of
Wolstencroft \& van Breda (\cite{WOL67}) and Ashburn (\cite{ASH54}).
Figure~\ref{fig:fracscatunit} shows the resulting scattering fractions for
Rayleigh and Mie scattering for different optical depths, which have been
converted into wavelengths based on the Cerro Paranal mean transmission curve
discussed in Sect.~\ref{sec:transmission}. For zenith distances that are
realistic for astronomical observations, wavelengths longer than about
$0.45$\,$\mu$m for Rayleigh scattering, and all wavelengths for aerosol
scattering, the scattering fractions stay below 10\% for a source of uniform
brightness. Scattering by air molecules and the ground surface can be important
for lines of sight close to the horizon and near-UV wavelengths.
Figure~\ref{fig:fracscatunit} indicates scattering fractions of about 40\%
under these conditions. Moreover, there appears to be a downturn of the
scattering fraction for very high zenith distances (see also Staude
\cite{STA75}), where the line-of-sight direction becomes nearly opaque. In this
case, the light that is not extinguished has to be scattered near the observer
and the incident photons have to originate close to the zenith.

The results of scattering calculations for more realistic source intensity
distributions will be discussed in Sects.~\ref{sec:stars}, \ref{sec:zodiac},
and \ref{sec:lines}.

\section{Moon, stars, and interplanetary dust}\label{sec:scatcomp}

Bright astronomical objects affect night-sky observations by the scattering of
their radiation into the line of sight. The relevant initial radiation sources
are the Sun and the summed-up light of all other stars. The solar radiation
must first be scattered by interplanetary dust grains and the Moon surface to
contribute to the night-sky brightness. In the following, we discuss these
scattering-related sky model components, i.e. scattered moonlight
(Sect.~\ref{sec:moon}), scattered starlight (Sect.~\ref{sec:stars}), and
zodiacal light (Sect.~\ref{sec:zodiac}). Another source of scattered light is
man-made light pollution, which can be neglected for Cerro Paranal. Patat
(\cite{PAT03}) did not detect spectroscopic signatures of such contamination.

\subsection{Scattered moonlight}\label{sec:moon}

\begin{figure}
\centering
\includegraphics[width=8.8cm,clip=true]{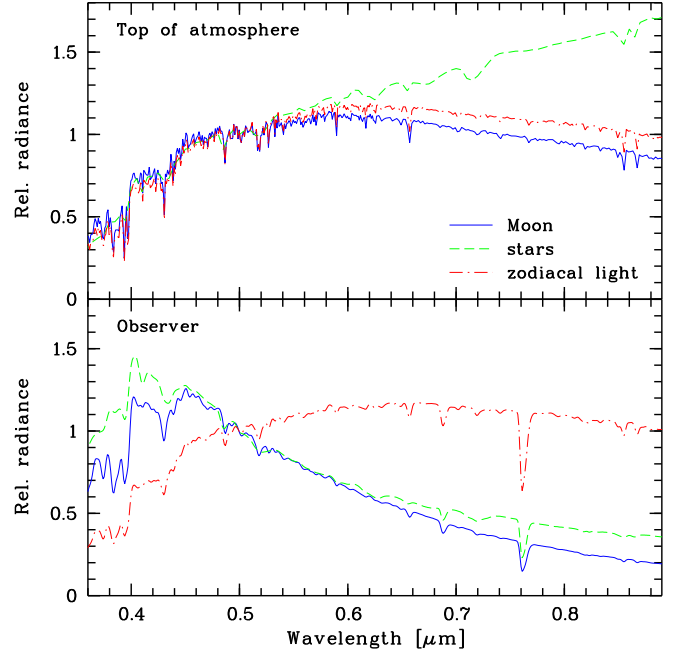}
\caption[]{Effects of atmospheric scattering on radiation originating outside
of the atmosphere. The displayed radiation sources are Moon (blue solid lines),
stars (green dashed lines), and interplanetary dust (red dash-dotted lines).
All spectra are normalised to unity at 0.5\,$\mu$m. In the {\em upper panel},
a solar spectrum is assumed as top-of-atmosphere spectrum for moonlight (see
Sect.~\ref{sec:moon}). A slightly reddened solar spectrum is taken for solar
radiation scattered at interplanetary dust grains (see Sect.~\ref{sec:zodiac}).
Integrated starlight tends to be redder than sunlight at long wavelengths (see
Sect.~\ref{sec:stars}). The {\em lower panel} shows the resulting
low-resolution spectra, after light has been scattered in the Earth's
atmosphere, for observing conditions as listed in Table~\ref{tab:modelpar}.
While the lunar and stellar contributions represent scattered radiation only,
the dust-related zodiacal light consists of scattered and direct light.}
\label{fig:compscatextrad}
\end{figure}

For observing dates close to Full Moon, scattered moonlight is by far the
brightest component of the optical night-sky background. In particular, blue
wavelengths are affected, since the Moon spectrum resembles a solar spectrum
and the efficency of Rayleigh/Mie scattering increases towards shorter
wavelengths (see Fig.~\ref{fig:transcurve}).

The Moon can be considered a point source. For this reason, it is not necessary
to perform the scattering calculations for extended sources, as discussed in
Sect.~\ref{sec:scattering}. Instead, a semi-analytical model for the
photometric $V$ band by Krisciunas \& Schaefer (\cite{KRI91}) is used for the
sky model, which has been extended into a spectroscopic version.

Based on Krisciunas \& Schaefer (\cite{KRI91}), we compute the Moon-related sky
surface brightness as the sum of the contributions from Rayleigh and Mie
scattering
\begin{equation}\label{eq:lunbrightsum}
B_\mathrm{moon}(\lambda) = B_{\mathrm{moon,R}}(\lambda) + B_{\mathrm{moon,M}}(\lambda),
\end{equation}
where
\begin{equation}\label{eq:lunbright}
B_\mathrm{moon,R/M}(\lambda) =
f_\mathrm{R/M}(\rho) \,I^{*}(\lambda) \,t^{\,X_\mathrm{moon}}(\lambda)
\,(1 - t_\mathrm{R/M}^{\,X_0}(\lambda)).
\end{equation}
The empirical scattering functions
\begin{equation}\label{eq:r-scatfunc}
f_\mathrm{R}(\rho) = 10^{\,5.70} \left(1.06 + \cos^2(\rho)\right)
\end{equation}
for Rayleigh scattering and
\begin{equation}\label{eq:m-scatfunc}
f_\mathrm{M}(\rho) = 10^{\,7.15 \,- \,(\rho / 40)}
\end{equation}
for Mie scattering depend on the angular separation of Moon and object $\rho$,
which is restricted to angles greater than $10\degr$. The functions deviate
from the ones of Krisciunas \& Schaefer (\cite{KRI91}) by factors of 2.2 and 10
respectively. These corrections are necessary because of the separation of the
two scattering processes in Eqs.~\ref{eq:lunbrightsum} and \ref{eq:lunbright},
the model extension in wavelength, and an optimisation for the observing
conditions at Cerro Paranal. The two correction factors were derived separately
from a comparison with optical VLT data (see Sect.~\ref{sec:data}) by using the
increasing importance of Rayleigh scattering with respect to Mie scattering for
shorter wavelengths and larger scattering angles $\rho$ (see
Sect.~\ref{sec:scattering}). The factor of 10 for Mie scattering is highly
uncertain due to a lack of clear constraints from the available data set. The
Moon illuminance is proportional to
\begin{equation}\label{eq:moonillu}
I^{*} \propto 10^{\,-\,0.4\,(0.026 \,|\phi| \,+ \,4.0 \,\times \,10^{-9} \,\phi^4)} \times
\left(d_\mathrm{moon}\right)^{-2},
\end{equation}
where the Moon distance $d_\mathrm{moon}$ can vary up to 5.5\% relative to the
mean value. The lunar phase angle $\alpha = 180\degr - \phi$ is the separation
angle of Moon and Sun as seen from Earth. It can be estimated from the
fractional lunar illumination (FLI) by
\begin{equation}\label{eq:alpha}
\alpha \approx \arccos(1 - 2 \,\mathrm{FLI}),
\end{equation}
since the influence of the lunar ecliptic latitude on $\alpha$ is negligible.
$B_{\mathrm{moon}}$ also depends on the atmospheric transmission for the airmass
of the Moon $t^{\,X_{\mathrm{moon}}}$, for which the Patat et al. (\cite{PAT11})
extinction curve is used (see Fig.~\ref{fig:transcurve}). Lastly, it depends
on a term that describes the amount of scattered moonlight in the viewing
direction, which can be roughly approximated by $1 - t_{\mathrm{R/M}}^{\,X_0}$ for
either Rayleigh or Mie scattering and the target airmass $X_0$. The airmasses
of the Krisciunas \& Schaefer (\cite{KRI91}) model are computed using
\begin{equation}\label{eq:scatterairmass}
X = \left(1 - 0.96 \sin^2(z)\right)^{-0.5},
\end{equation}
which is especially suited for scattered light, because of a low limiting
airmass at the horizon (cf. Eq.~\ref{eq:airmass}).

The Krisciunas \& Schaefer model was developed only for the $V$ band. It can
be extended to other wavelength ranges by using the Cerro Paranal extinction
curve (see Sect.~\ref{sec:transmission}) and its Rayleigh and Mie components
for the transmission curves $t$, $t_{\mathrm{R}}$, and $t_{\mathrm{M}}$ in
Eq.~\ref{eq:lunbright} and by assuming that the Moon spectrum resembles the
spectrum of the Sun (see Fig.~\ref{fig:compscatextrad}). The latter is modelled
by the solar spectrum of Colina et al. (\cite{COL96}) scaled to the $V$-band
brightness of the moonlight scattering model. For most situations, this
approach results in systematic uncertainties of no more than 10\% of the total
night-sky flux compared to observations. The statistical uncertainties are in
the order of 10 to 20\% (see Sect.~\ref{sec:discussion}). The errors become
larger for Full Moon and/or target positions close to the Moon
($\rho \la 30\degr$). The latter especially concerns red wavelengths, where the
contribution of Mie forward scattering to the total intensity of scattered
light is particularly high. Consequently, the strongly varying aerosol
properties have to be known (which is a challenging task) to allow for a good
agreement of model and observations. However, optical astronomical observations
close to the Moon are unlikely. Therefore, this is not a critical issue for the
sky model.

\subsection{Scattered starlight}\label{sec:stars}

Like moonlight, starlight is also scattered in the Earth's atmosphere. However,
stars are distributed over the entire sky with a distribution maximum towards
the centre of the Milky Way. This distribution requires the use of the
scattering model for extended sources described in Sect.~\ref{sec:scattering}.
Since scattered starlight is only a minor component compared to scattered
moonlight and zodiacal light (see Fig.~\ref{fig:logcomp}), it is sufficient
for an ETC application to compute a mean spectrum. This simplification allows
us to avoid the introduction of sky model parameters such as the galactic
coordinates, which have a very low impact on the total model flux.

For the integrated starlight (ISL), we use Pioneer\,10 data at 0.44\,$\mu$m
(Toller \cite{TOL81}; Toller et al. \cite{TOL87}; Leinert et al.
\cite{LEI98}). These data are almost unaffected by zodiacal light. The small
``hole'' in the data set towards the Sun has been filled by interpolation.
Since the Pioneer\,10 maps do not include stars brighter than 6.5\,mag in $V$,
a global correction given by Melchior et al. (\cite{MEL07}) is applied, which
increases the total flux by about 16\%. Scattering calculations are only
performed for the distribution of starlight in the $B$ band, i.e. the
dependence of this distribution on wavelength is neglected. This approach is
supported by calculations of Bernstein et al. (\cite{BER02}), indicating that
the shapes of ISL spectra are very stable, except for regions in the dustiest
parts of the Milky Way plane. The wavelength dependence of the scattered
starlight is considered by multiplying the representative ISL mean spectrum of
Mattila (\cite{MAT80}) by the resulting wavelength-dependent amount of
scattered light (for illustration see Figs.~\ref{fig:fracscatunit} and
\ref{fig:compscatextrad}). Since the mean ISL spectrum of Mattila only
covers wavelengths up to 1\,$\mu$m, we extrapolated the spectral range by
fitting an average of typical spectral energy distributions (SED) for early-
and late-type galaxies produced by the SED-fitting code CIGALE (see Noll et al.
\cite{NOL09}). Since the SED slopes are very similar in the near-IR, the choice
of the spectral type is not crucial. The solar and the mean ISL spectrum
are similar at blue wavelengths, but the ISL spectrum has a redder slope at
longer wavelengths (see Fig.~\ref{fig:compscatextrad}). The differences
illustrate the importance of K and M stars for the ISL.

For the desired average scattered light spectrum, we run our scattering code
for different combinations of zenithal optical depth, zenith distance, azimuth,
and sidereal time for Rayleigh and Mie scattering (see
Sect.~\ref{sec:scattering}). The step sizes for the latter three parameters
were $10\degr$, $45\degr$, and 2\,h, respectively. Zenith distances were only
calculated up to $50\degr$, since this results in a mean airmass of about
$1.25$ for the covered solid angle, which is typical of astronomical
observations. The mean scattering intensities for each optical depth $\tau_0$
were then computed. This was translated into a spectrum by using the relation
between $\tau_0$ and wavelength as provided by the Cerro Paranal extinction
curve (see Sect.~\ref{sec:transmission}). Finally, the mean ISL spectrum
(normalised to the $B$-band flux) was multiplied. The sky model code does not
change the final spectrum, except for the molecular absorption, which is
adjusted depending on the bimonthly period (see Sect.~\ref{sec:transmission}).
For the effective absorption airmass, the mean value of $1.25$ is assumed.

The resulting spectrum shows an intensity of about 13\,\radunits{} at
0.4\,$\mu$m. At $0.6$\,$\mu$m, there is only half of this intensity (see
Fig.~\ref{fig:compscatextrad}). The results are in good agreement with the
findings of Bernstein et al. (\cite{BER02}).

\subsection{Zodiacal light}\label{sec:zodiac}

Zodiacal light is caused by scattered sunlight from interplanetary dust grains
in the plane of the ecliptic. A strong contribution is found for low
absolute values of ecliptic latitude $\beta$ and heliocentric ecliptic
longitude $\lambda - \lambda_{\sun}$. The brightness distribution provided by
Levasseur-Regourd \& Dumont (\cite{LEV80}) and Leinert et al. (\cite{LEI98})
for $0.5$\,$\mu$m shows a relatively smooth decrease for increasing elongation,
i.e. angular separation of object and Sun. A striking exception is the local
maximum of the so-called gegenschein at the antisolar point in the ecliptic.
The spectrum of the optical zodical light is similar to the solar spectrum
(Colina et al. \cite{COL96}), but slightly reddened (see
Fig.~\ref{fig:compscatextrad}). We apply the relations given in Leinert et al.
(\cite{LEI98}) to account for the reddening. The correction is larger for
smaller elongations. Thermal emission of interplanetary dust grains in the IR
is neglected in the sky model, since the airglow components of atmospheric
origin (see Fig.~\ref{fig:logcomp}) completely outshine it. In the optical,
zodiacal light is a significant component of the sky model. A contribution of
about 50\% is typical of the $B$ and $V$ bands when the Moon is down (see
Sect.~\ref{sec:photo}). At longer wavelengths, the fraction decreases due to
the increasing importance of the airglow continuum (see Sect.~\ref{sec:cont}).

The model of the zodiacal light presented in Leinert et al. (\cite{LEI98})
describes the characteristics of this emission component outside of the
Earth's atmosphere. Ground-based observations of the zodical light also have to
take atmospheric extinction into account. Since zodiacal light is an extended
radiation source, the observed intensity is a combination of the extinguished
top-of-atmosphere emission in the viewing direction and the intensity of light
scattered into the line of sight (see Fig.~\ref{fig:compscatextrad}). The
latter can be treated by scattering calculations as discussed in
Sect.~\ref{sec:scattering}. Scattering out of and into the line of sight leads
to an effective extinction and this can be expressed by an effective optical
depth
\begin{equation}\label{eq:taueff}
\tau_\mathrm{eff} = \tau_\mathrm{0,eff} \,X = f_\mathrm{ext} \,\tau_0 \,X
\end{equation}
(cf. Eq.~\ref{eq:extinction}; see also Bernstein et al. \cite{BER02}).
Consequently, the scattering properties of the zodiacal light can be described
by the factor $f_\mathrm{ext}$ alone. This parameter shows only a weak
dependence on $\tau_0$ and, hence, wavelength. For Rayleigh scattering in the
range from $U$ to $I$ band, we find an uncertainty of only about 4\%. For Mie
scattering from $V$ to $J$, a similar variation is found. We obtain this result
by calculating $f_\mathrm{ext}$ for a grid of optical depths, zenith distances,
azimuths, sideral times, and solar ecliptic longitudes. We take the same grid
as for scattered starlight (see Sect.~\ref{sec:stars}) plus solar ecliptic
longitudes in steps of $90\degr$, i.e. for each of the four seasons, one data
set was calculated. Scattering calculations were only carried out for solar
zenith distances of at least $108\degr$. This restriction excludes daytime and
twilight conditions. As for scattered starlight, we consider target zenith
distances $z$ up to $50\degr$ (see Sect.~\ref{sec:stars}). For rare high
zodiacal light intensities, we also let $50\degr < z \le 70\degr$. The $z$
limits do not significantly affect $f_\mathrm{ext}$. A test with
$0\degr \le z \le 70\degr$ showed a variation of $f_\mathrm{ext}$ in the order
of a few per cent only.

\begin{figure}
\centering
\includegraphics[width=8.8cm,clip=true]{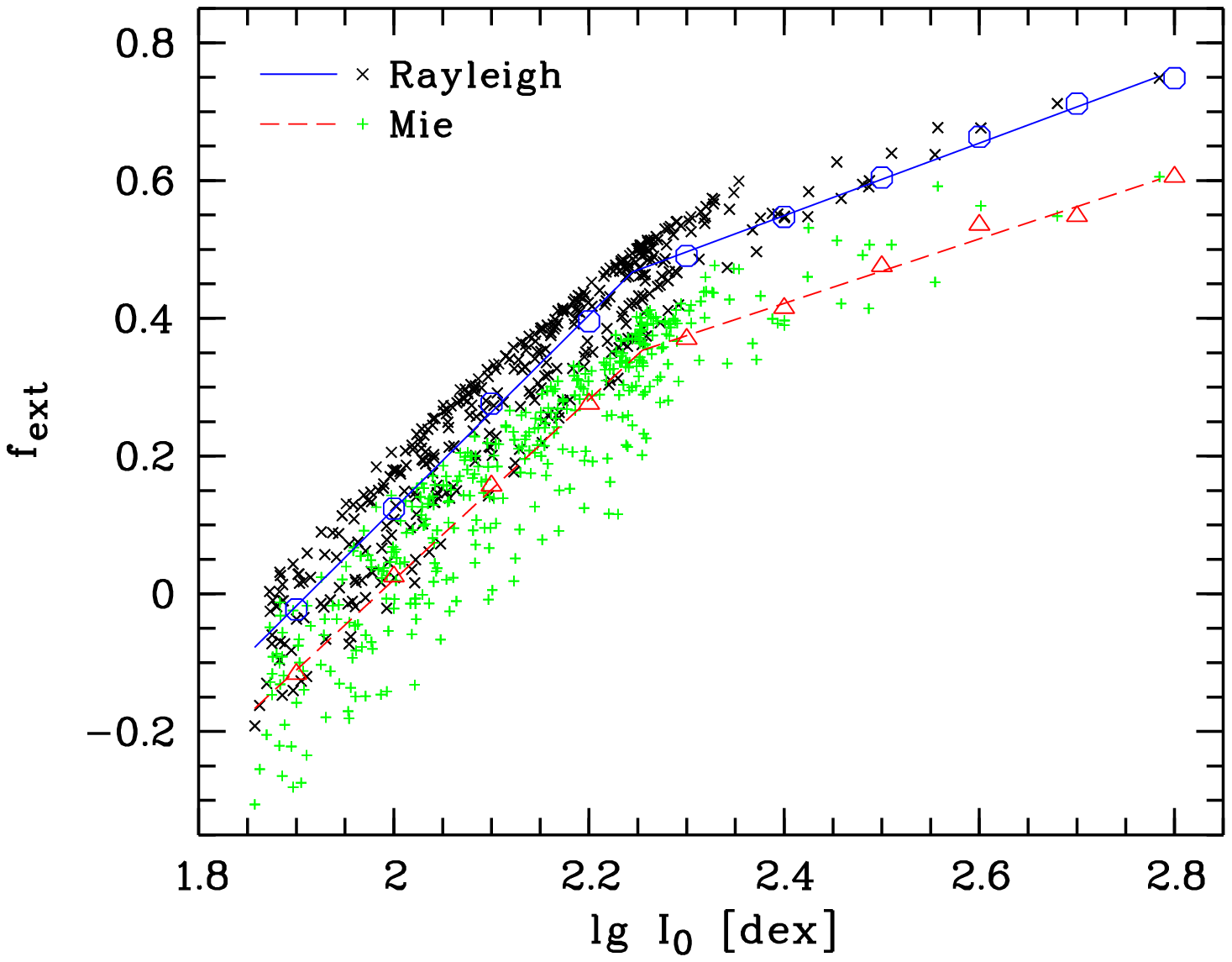}
\caption[]{Extinction reduction factor $f_{\mathrm{ext}}$ for zodiacal light as a
function of the line-of-sight top-of-atmosphere intensity $I_0$ of the zodiacal
light in dex. The intensities correpond to those given in Table~17 of Leinert
et al. (\cite{LEI98}), i.e. the unit is
$10^{-8}\,\mathrm{W\,m}^{-2}\,\mu\mathrm{m}^{-1}\,\mathrm{sr}^{-1}$
($= 1.690$\,\radunits{} at 0.5\,$\mu$m). The factors
were calculated for different ecliptic longitudes of the Sun, sidereal times,
and line-of-sight zenith distances (up to 50$\degr$ in general and up to
70$\degr$ for $\log I_0 > 2.4$) and azimuths. Only night-time data points with
solar zenith distances greater than 108$\degr$ are considered. For Rayleigh
scattering (black 'x' symbols), a zenithal optical depth of 0.27 is taken,
corresponding to a wavelength of 0.4\,$\mu$m for the Cerro Paranal extinction
curve (see Sect.~\ref{sec:transmission}). For Mie scattering (green '+'
symbols), $\tau_0$ is fixed to 0.01, a value reached at about 1.2\,$\mu$m. The
figure also shows average $f_{\mathrm{ext}}$ in 0.1\,dex $I_0$ bins for both
scattering modes (Rayleigh: blue circles; Mie: red triangles). The resulting
fits based on these average values are displayed by solid (Rayleigh) and dashed
lines (Mie).}
\label{fig:fiteffzodiac}
\end{figure}

\begin{figure}
\centering
\includegraphics[width=8.8cm,clip=true]{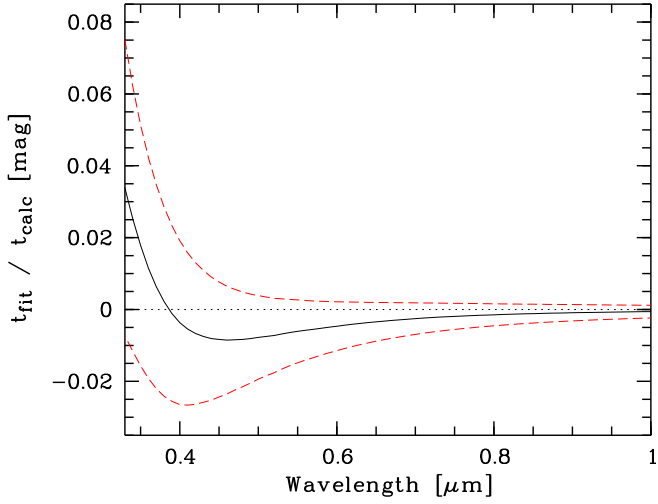}
\caption[]{Ratio of the zodiacal light transmissions of the fits as shown in
Fig.~\ref{fig:fiteffzodiac} and the full scattering calculations in magnitudes.
The wavelength-dependent mean values and 1\,$\sigma$ deviations are displayed
as solid and dashed lines respectively.}
\label{fig:qualzodiacscat}
\end{figure}

For an efficient implementation in the ETC, we searched for a parametrisation
of $f_\mathrm{ext}$ that simplifies the treatment of scattering and minimises
the number of required input parameters. We found that the intial intensity of
the zodiacal light in the viewing direction $I_0$ indicates the tightest
relation with $f_\mathrm{ext}$. The next-best parameter is the ecliptic
latitude. Figure~\ref{fig:fiteffzodiac} shows $f_\mathrm{ext}$ versus the
logarithm of $I_0$ as provided by Leinert et al. (\cite{LEI98}) for Rayleigh
and Mie scattering. Since the variation of $f_\mathrm{ext}$ with $\tau_0$ can be
neglected, the optical depth is fixed to $0.27$ for Rayleigh and $0.01$ for Mie
scattering. These values correspond to the wavelengths $0.4$ and $1.2$\,$\mu$m
respectively, which represent typical wavelengths dominated by the two
different scattering processes (see Fig.~\ref{fig:transcurve}). The
distribution of data points in Fig.~\ref{fig:fiteffzodiac} exhibits a clear
increase of $f_\mathrm{ext}$ with $I_0$. The slopes appear to be very similar
for both scattering processes. The extinction reduction factor
$f_\mathrm{ext,M}$ for Mie scattering tends to be about 0.1 lower than
$f_\mathrm{ext,R}$ for Rayleigh scattering. The difference increases with
increasing $I_0$. For very low zodiacal light intensities in the viewing
direction, $f_\mathrm{ext}$ becomes negative. Here, the scattering of light from
directions of high zodiacal light brightness into the line of sight
overcompensates the extinction. For viewing directions with higher $I_0$,
this process is less effective and causes higher $f_\mathrm{ext}$. The
investigated parameter space does not indicate values higher than about 0.75,
i.e. the zodiacal light in the direction of astronomical observing targets is
always significantly less attenuated in the Earth's atmosphere than a point
source (cf. Leinert et al. \cite{LEI98}).

The good correlation of $f_\mathrm{ext}$ and $I_0$ allows us to describe the
reduction of the optical depth with $I_0$ only. The unextinguished zodiacal
light is derived from the ecliptic coordinates, which are given sky model
parameters (see Table~\ref{tab:modelpar}). Figure~\ref{fig:fiteffzodiac}
displays the resulting linear fits to the data based on average $f_\mathrm{ext}$
for 0.1\,dex bins of $\log I_0$. The parameters obtained by these split fits
are
\begin{equation}\label{eq:r-zodiacfit}
f_\mathrm{ext,R} =
\begin{cases}
1.407 \log I_0 - 2.692 & \text{if } \log I_0 \le 2.244, \\
0.527 \log I_0 - 0.715 & \text{if } \log I_0 > 2.244 \\
\end{cases}
\end{equation}
for Rayleigh scattering and
\begin{equation}\label{eq:m-zodiacfit}
f_\mathrm{ext,M} =
\begin{cases}
1.309 \log I_0 - 2.598 & \text{if } \log I_0 \le 2.255, \\
0.468 \log I_0 - 0.702 & \text{if } \log I_0 > 2.255 \\
\end{cases}
\end{equation}
for Mie scattering with $I_0$ in
$10^{-8}\,\mathrm{W\,m}^{-2}\,\mu\mathrm{m}^{-1}\,\mathrm{sr}^{-1}$. For the fits
of the lower $I_0$, the $f_\mathrm{ext}$ standard deviations of the data from
the regression lines are 0.06 and 0.07 for Rayleigh and Mie scattering
respectively. For the higher $I_0$, systematic uncertainties are probably
higher than statistical variations, since these values are rare when
$z \le 70\degr$ and $z_{\,\sun} \ge 108\degr$.

The quality of the parametrisation in Eqs.~\ref{eq:r-zodiacfit} and
\ref{eq:m-zodiacfit} is illustrated by Fig.~\ref{fig:qualzodiacscat}, which
shows the average ratio of the fitted and originally calculated transmission
for the entire parameter grid as derived from Eqs.~\ref{eq:extinction} and
\ref{eq:taueff} in magnitudes. In almost the entire wavelength range, the
accuracy of the mean is better than 1\%. Only at very short wavelengths, the
systematic errors become higher. The intersection of the zero line at about
$0.4$\,$\mu$m is caused by the selection of $\tau_0 = 0.27$ for the
$f_\mathrm{ext,R}$ fitting. The $1\sigma$ range suggests uncertainties of no
more than 3 to 4\% for most wavelengths. These results suggest that the
parametrisation of $\tau_\mathrm{eff}$ does not cause higher errors than those
expected from the uncertainties of the scattering calculations (see
Sect.~\ref{sec:scattering}). The total errors from the simple treatment of
multiple scattering and ground reflection, neglection of polarisation,
simplified particle distributions, and uncertainties in the aerosol scattering
parameters should be in the order of 10\% of $I_\mathrm{scat}$. This translates
into uncertainties in the transmission of a few per cent at blue wavelengths,
which is somewhat higher than the uncertainties from the $f_\mathrm{ext}$ fit.

Finally, the reduction of the zodiacal light intensity by molecular absorption
has to be considered. Strong absorption is only important for relatively long
wavelengths, where scattering affects only a few per cent of the incoming light
(see Fig.~\ref{fig:transcurve}), assuming zenith distances of typical
astronomical observations. Hence, the direct component in the viewing direction
is usually much brighter than the scattered light from other directions. For
this reason, we can safely assume that the target airmass is well suited to
derive the transmission reduction by molecular absorption. Differences between
target and effective scattering airmasses can be neglected.

\section{Airglow emission}\label{sec:airglow}

Atmospheric emission at wavelengths from the near-UV to the near-IR originates
in the upper atmosphere, especially in the mesopause region at an altitude of
about 90\,km and in the ionospheric F2 layer at about 270\,km. In contrast to
the thermal radiation in the mid-IR by atmospheric greenhouse gases in the
lower atmosphere, this so-called airglow (see Khomich et al. \cite{KHO08} for a
comprehensive discussion) is caused by a highly non-LTE\footnote{local
thermodynamic equilibrium} emission process called chemiluminescence, i.e. by
chemical reactions that lead to light emission by the decay of excited
electronic states of reaction products. Consequently, atmospheric radiative
transfer codes like LBLRTM (Clough et al. \cite{CLO05}; see also
Sect.~\ref{sec:transmission}) cannot be used to calculate this emission.
Moreover, theoretical analysis is very challenging because of the high
variability of airglow on various time scales. The complex reactions are
affected by the varying densities of the involved molecules, atoms, and ions in
different excitation states. Therefore, airglow can best be treated by a
semi-empirical approach. The section starts with a general discussion of
airglow extinction (Sect.~\ref{sec:airglowext}). In Sect.~\ref{sec:data}, we
present the data set that was used to derive the model. Detailed discussions
of the line and continuum components of the model can be found in
Sects.~\ref{sec:lines} and \ref{sec:cont} respectively.

\subsection{Extinction of airglow emission}\label{sec:airglowext}

\begin{figure}
\centering
\includegraphics[width=8.8cm,clip=true]{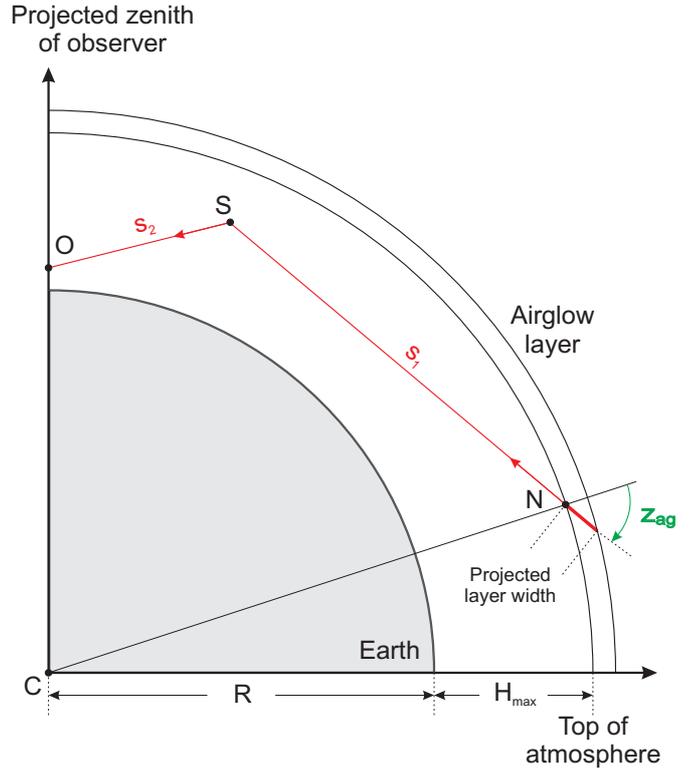}
\caption[]{Illustration of the projected emission layer width used for the
derivation of the beam-specific, incident airglow intensities in the 3D
scattering calculations. The extension of the scattering atmosphere is limited
to the lower boundary of the airglow emission layer. Point $O$ can be in a
different plane from the other points (cf. Fig.~\ref{fig:scatsketch}).}
\label{fig:agscatsketch}
\end{figure}

\begin{figure}
\centering
\includegraphics[width=8.8cm,clip=true]{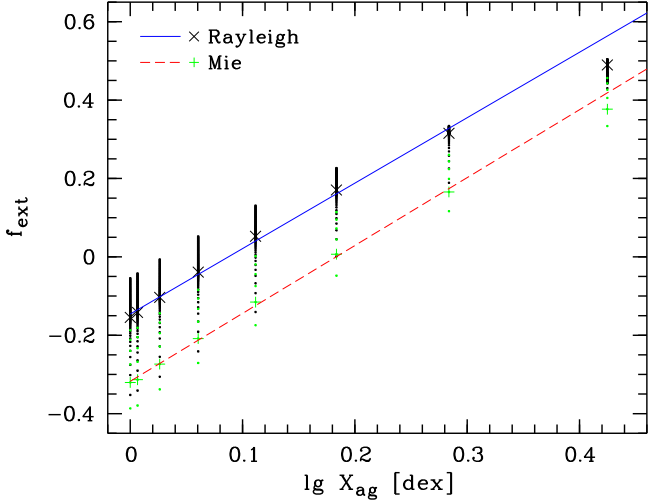}
\caption[]{Extinction reduction factor $f_{\mathrm{ext}}$ for airglow as a
function of airmass (see Eq.~\ref{eq:airglowairmass}) in dex. The factors were
calculated for different line-of-sight zenith distances $z$ (in 10$\degr$
steps) and zenithal optical depths $\tau_0$. For Rayleigh ($\tau_0 \le 0.99$)
and Mie scattering ($\tau_0 \le 0.05$), the values for $\tau_0 = 0.27$ and
$0.01$ are marked by black 'x' and green '+' symbols respectively. For
$z \le 60\degr$, these values were used to derive airmass-dependent relations
for $f_{\mathrm{ext}}$. The resulting fits are displayed by blue solid (Rayleigh)
and red dashed lines (Mie).}
\label{fig:fiteffairglow}
\end{figure}

\begin{figure}
\centering
\includegraphics[width=8.8cm,clip=true]{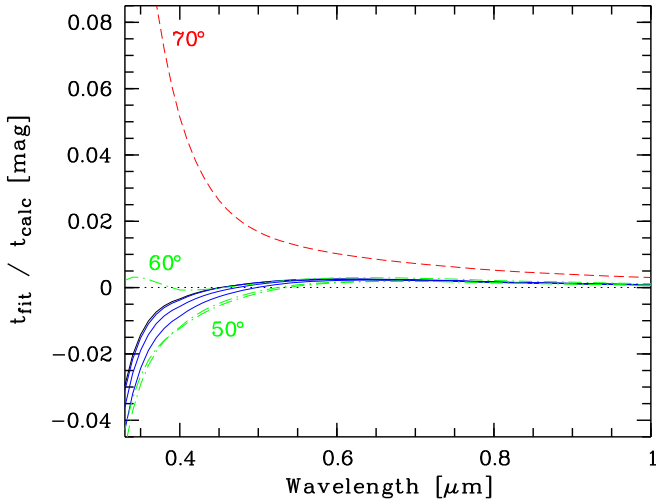}
\caption[]{Ratio of the airglow transmissions of the fits as shown in
Fig.~\ref{fig:fiteffairglow} and the full scattering calculations in
magnitudes. The wavelength-dependent results for line-of-sight zenith distances
from 0$\degr$ (black solid line) to 70$\degr$ (red dashed line) in 10$\degr$
steps are exhibited.}
\label{fig:qualairglowscat}
\end{figure}

The intensities of airglow lines and continuum rise with growing zenith
distance by the increase of the projected emission layer width. This behaviour
is expressed by the van Rhijn function
\begin{equation}\label{eq:vanrhijn}
\frac{I(z)}{I(0)} =
\left(1 - \left(\frac{R \sin(z)}{R + h}\right)^2\right)^{-0.5}
\end{equation}
(van Rhijn \cite{RHI21}). Here, $z$, $R$, and $h$ are the zenith distance, the
Earth's radius, and the height of the emitting layer above the Earth's
surface (see Table~\ref{tab:vardata}), respectively. The airglow intensity is
also affected by scattering and absorption in the lower atmosphere. Since
the airglow emission is distributed over the entire sky, we can derive the
effective extinction by means of scattering calculations as described in
Sect.~\ref{sec:scattering}. Some modifications have to be applied, since
airglow emission originates in the atmosphere itself and because of the van
Rhijn effect. The top of atmosphere is reduced from the default
$H_\mathrm{max} = 200$\,km to 90\,km, which is assumed to be representative of
airglow emission layers. The few atomic lines originating in the ionospheric
F2 layer are neglected here (see Sect.~\ref{sec:lines}). The cut at the lower
height is not critical, since the mass of the excluded range is negligible
compared to the total mass of the atmosphere (see Eq.~\ref{eq:baroform}). The
projected emission layer width (see Fig.~\ref{fig:agscatsketch}) is considered
by calculating the zenith distance $z_\mathrm{\,ag}$ of each incoming beam as
seen from the corresponding entry point $N$ at $H_\mathrm{max}$. For our set-up,
the resulting beam-specific airglow intensity can be approximated by
\begin{equation}
I(z_\mathrm{\,ag}) = \frac{I_0}{\cos(z_\mathrm{\,ag})},
\end{equation}
since the centres of the scattering path elements $S$ are always located well
below the airglow emission layer height, which avoids
$z_\mathrm{\,ag} = 90\degr$. The intensity $I_0$ for a perpendicular incidence
at $N$ is assumed to be constant for the entire global airglow layer. The
scattering calculations only depend on the zenithal optical depth and the
target zenith distance.

Like for the zodiacal light (see Sect.~\ref{sec:zodiac}), we can derive an
optical depth reduction factor $f_\mathrm{ext}$ (Eq.~\ref{eq:taueff}) to
describe the airglow extinction properties. The representative optical depths
are again 0.27 for Rayleigh scattering (6\% uncertainty from $U$ to $I$ band)
and 0.01 for Mie scattering (4\% variation from $V$ to $J$ band). The only
remaining free parameter for $f_\mathrm{ext}$ is the zenith distance or airmass.
We take
\begin{equation}\label{eq:airglowairmass}
X_\mathrm{\,ag} = \left(1 - 0.972 \sin^2(z)\right)^{-0.5}
\end{equation}
(cf. Eq.~\ref{eq:scatterairmass}), which corresponds to the van Rhijn formula
(Eq.~\ref{eq:vanrhijn}) for the assumed airglow layer height of 90\,km.
$X_\mathrm{\,ag}$ is better suited for fitting $f_\mathrm{ext}$ and
Eq.~\ref{eq:taueff} than the Rozenberg (\cite{ROZ66}) airmass $X$
(Eq.~\ref{eq:airmass}) used for the zodiacal light (see
Sect.~\ref{sec:zodiac}), where the $z$ dependence of $f_\mathrm{ext}$ was
negligible. Figure~\ref{fig:fiteffairglow} shows the extinction reduction
factor as a function of the logarithm of $X_\mathrm{\,ag}$. There is a strong
increase in $f_\mathrm{ext}$ with airmass, and the factors for the Mie
scattering tend to be about $0.15$ lower than those for Rayleigh scattering.
These trends are in good agreement with our findings for the zodiacal light.
However, the typical values of $f_\mathrm{ext}$ are significantly lower. For
most zenith distances relevant for astronomical observations, the factors are
close to zero. Consequently, the effective extinction of airglow is very low
(see Chamberlain \cite{CHA61}). For observations close to zenith, there is even
a significant brightening of the airglow emission by the scattering into the
line of sight from light with a long path length through the emission layer.
Therefore, using a point-source transmission curve for airglow extinction
correction will provide worse results than having no correction. As indicated
by Fig.~\ref{fig:fiteffairglow}, a rough correction of airglow emission using
the van Rhijn equation is reasonable up to zenith distances of $50\degr$ to
$60\degr$, depending on the ratio of Rayleigh and Mie scattering. At higher
zenith distances, the compensation of extinction of direct light by scattered
light from other directions becomes less efficient. At the largest $z$, the
intensity can be more reduced by extinction than enhanced by the van Rhijn
effect. For this reason, there is a zenith distance with a maximum airglow
intensity, which depends on the optical depth and, hence, the wavelength. For
the Cerro Paranal extinction curve (see Sect.~\ref{sec:transmission}), we found
a $z_\mathrm{max}$ of about $70\degr$ at $0.35$\,$\mu$m and $85\degr$ at
$0.7$\,$\mu$m.

For the sky model, $f_\mathrm{ext}$ was fit up to a zenith distance of
$60\degr$, where $f_\mathrm{ext}$ shows an almost linear relation with the
logarithm of $X_\mathrm{\,ag}$ (see Fig.~\ref{fig:fiteffairglow}). A worse
agreement at large zenith distances is tolerable, since astronomical
observations are rarely carried out at those angles. The parameters obtained by
the fits are
\begin{equation}\label{eq:r-airglowfit}
f_\mathrm{ext,R} = 1.669 \log X_\mathrm{\,ag} - 0.146
\end{equation}
for Rayleigh scattering and
\begin{equation}\label{eq:m-airglowfit}
f_\mathrm{ext,M} = 1.732 \log X_\mathrm{\,ag} - 0.318
\end{equation}
for Mie scattering. For the fixed optical depths and the fitted $z$ range, the
fit uncertainties of $f_\mathrm{ext}$ are only about 0.01.
Figure~\ref{fig:qualairglowscat} displays the accuracy of the fits as a
function of wavelength. For this plot, the Cerro Paranal extinction curve (see
Sect.~\ref{sec:transmission}) was used to convert optical depths into
wavelengths. For the fitted range of zenith distances, the deviations between
fitted and the true transmission curves are below 1\%, at least for wavelengths
longer than $0.4$\,$\mu$m. As expected, the deviations for $z = 70\degr$ are
significantly larger than for the smaller zenith distances. Nevertheless,
for most prominent airglow lines (see Sect.~\ref{sec:lines}), the errors are
below 2\%, which also makes the parametrisation sufficient for large zenith
distances.

Apart from scattering, molecular absorption in the lower atmosphere affects
the observed airglow intensity. However, in the optical these absorptions are
usually small (see Fig.~\ref{fig:transcurve}). Exceptions are the airglow
O$_2$(b-X)(0-0) and O$_2$(b-X)(1-0) bands at 762 and 688\,nm, which suffer from
heavy self-absorption at low altitudes, and can be observed as absorption bands
only. As discussed in Sect.~\ref{sec:zodiac}, the effective molecular
absorption can be well approximated by taking the transmission spectrum for the
target zenith distance. For the airglow continuum component,
this is easily applied. For the line component, this is challenging, since
airglow emission as well as telluric absorption lines are very narrow.
Therefore, an effective absorption for each airglow line has to be derived at
very high resolution. Resolving the airglow lines requires resolutions in the
order of $10^6$. The absorption lines tend to be broader than the emission
lines. LBLRTM (see Clough et al. \cite{CLO05}) easily produces spectra with the
required resolution. For the airglow emission, we use a line list with good
wavelength accuracy, as discussed in Sect.~\ref{sec:lines}. In the upper
atmosphere, broadening of lines by collisions can be neglected due to the low
density (see Khomich et al. \cite{KHO08}). Thus, the shape and width of airglow
lines are determined by Doppler broadening, which can be derived by the
molecular weight of the species and ambient temperature. The latter is in the
order of 200\,K in the mesopause region, which explains the low Doppler widths.
For example, the FWHM of an OH line (see Sect.~\ref{sec:lines}) at 0.8\,$\mu$m
is about 2\,picometers (pm). The few lines originating in the thermospheric
ionosphere experience distinctly higher temperatures of about 1000\,K. The line
absorption correction is applied as the multiplication of the initial line
intensities by the suitable line transmissions from a pre-calculated list,
which is derived from the library transmission spectrum corresponding to the
selected observing conditions (see Sect.~\ref{sec:transmission}).

The airglow line absorptions show significant deviations from the continuum
absorption at similar wavelengths. As already mentioned, strong self-absorption
is found for the O$_2$ ground state transitions in the optical. However, most
other airglow lines tend to be less absorbed than the continuum. In particular,
the H$_2$O absorption bands (see Sect.~\ref{sec:transmission}) have little
effect. This can be explained by the relatively low width of the telluric lines
compared to the typical distance between strong absorption lines. Hence, the
chance to find an airglow line at the centre of a strong absorption feature is
relatively low.

\subsection{Data set for airglow analysis}\label{sec:data}

\begin{figure}
\centering
\includegraphics[width=8.8cm,clip=true]{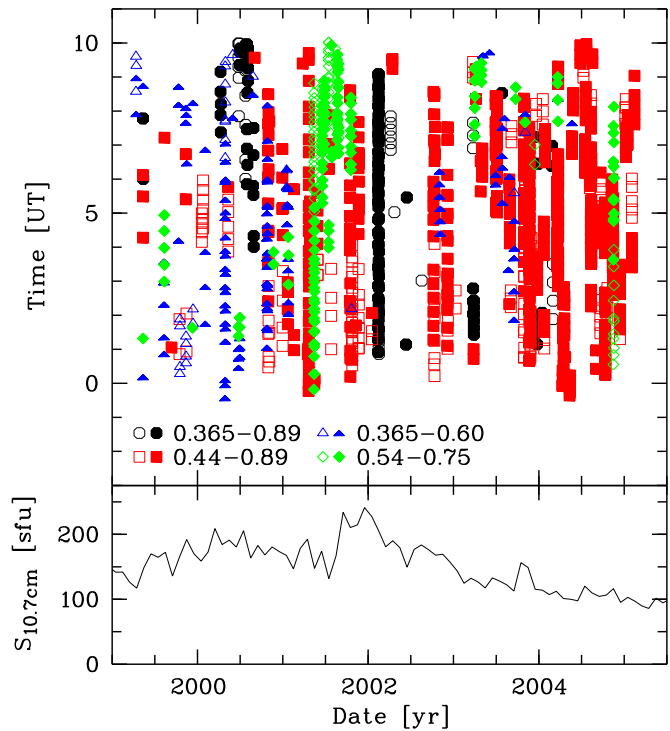}
\caption[]{Date, time, and solar activity for the VLT FORS\,1 observations of
the Patat (\cite{PAT08}) spectroscopic data set. The solar activity is given by
the solar radio flux measured at 10.7\,cm in sfu. The FORS\,1 spectra were
taken with different instrument set-ups, which are indicated by different
symbols and colours. The plot legend identifies the different set-ups by their
covered wavelength ranges in $\mu$m. Open symbols indicate that the data were
taken with the Moon above the horizon, whereas filled symbols refer to
dark-time observations.}
\label{fig:forssample}
\end{figure}

The airglow analysis (see Sects.~\ref{sec:lines} and \ref{sec:cont}) and the
sky model verification (see Sect.~\ref{sec:discussion}) were carried out
using a sample of 1186 FORS\,1 long-slit sky spectra (see
Fig.~\ref{fig:forssample}). The FORS\,1 data were collected from the ESO
archive and reduced by Patat (\cite{PAT08})\footnote{We excluded three spectra
from the original sample of 1189 spectra because of unreliable continua
concerning strength and shape.}. The spectra were taken with the
low/intermediate resolution grisms 600B (12\%), 600R (17\%), and 300V, the
latter with (57\%) and without (14\%) the order separation filter GG435. The
set-ups cover the wavelength range from 0.365 to 0.89\,$\mu$m. Wavelengths
below 0.44\,$\mu$m are the least covered, since only 600B and 300V spectra
without GG435 can be used there. The FORS\,1 spectra were taken between April
1999 and February 2005, i.e. observations during a phase of low solar activity
are not part of the present data set (see Fig.~\ref{fig:forssample}). The mean
and standard deviation of the Penticton-Ottawa solar radio flux at
10.7\,cm\footnote{\texttt{ftp://ftp.geolab.nrcan.gc.ca/data/solar\_flux/}}
(Covington \cite{COV69}) are 153 and 35 solar flux units (sfu = 0.01\,MJy)
respectively. The data set is also characterised by a mean zenith distance of
$31\degr$. The corresponding standard deviation is only $13\degr$ and the
maximum zenith distance is $67\degr$. The fraction of exposures with the Moon
above the horizon amounts to 26\%. 29 spectra were taken at sky positions
with Moon distances below $30\degr$ (see Sect.~\ref{sec:moon}). However,
exposures affected by strong zodical light contribution are almost completely
absent.

The FORS\,1 night-sky spectra were not extinction corrected and the response
curves for the flux calibration were derived by the Cerro Tololo standard
extinction curve (Stone \& Baldwin \cite{STO83}; Baldwin \& Stone
\cite{BALD84}). We corrected all the spectra to the more recent and adequate
Cerro Paranal extinction curve of Patat et al. (\cite{PAT11}, see
Sect.~\ref{sec:transmission}), assuming a mean airmass of 1.25 for the
spectrophotometric standard stars. This procedure resulted in corrections of
$+6$\% at the blue end to $-1$\% at the red end of the covered wavelength
range.

\subsection{Airglow lines}\label{sec:lines}

In the following, we treat airglow lines in detail. At first, the origin of the
airglow line emission, the classification of lines, and a derivation of a line
list for the sky model are discussed (Sect.~\ref{sec:linelist}). In the second
part, we investigate the variability of airglow lines
(Sect.~\ref{sec:linevar}).

\subsubsection{Line classes and line list}\label{sec:linelist}

\begin{figure}
\centering
\includegraphics[width=8.8cm,clip=true]{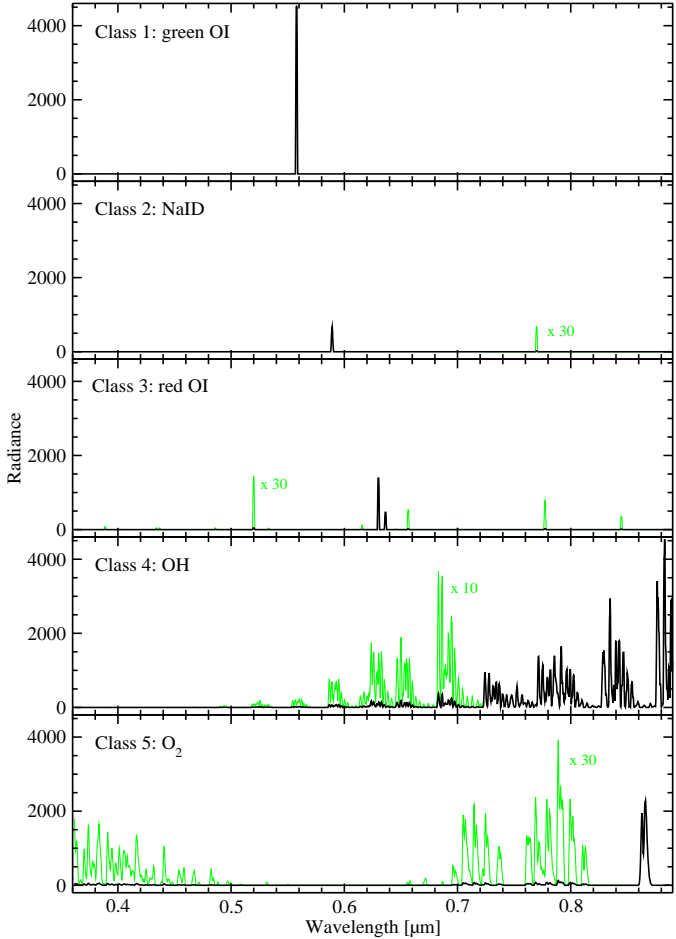}
\caption[]{Variability classes for airglow emission lines. The following groups
are defined: (1) green O\,I, (2) Na\,I\,D, (3) red O\,I, (4) OH, and (5) O$_2$.
The weak lines (green curves) are scaled by a factor of 30 for Na\,I\,D, red
O\,I, and O$_2$, and a factor of 10 for OH.}
\label{fig:varclasses}
\end{figure}

\begin{table*}
\caption[]{Basic properties of the Paranal airglow line and continuum
(0.543\,$\mu$m) variability as derived from the Patat (\cite{PAT08}) spectral
data set.}
\label{tab:vardata}
\centering
\vspace{5pt}
\begin{tabular}{c l c c c c c c}
\hline\hline
\noalign{\smallskip}
Property & Unit & [O\,I]\,5577 & Na\,I\,D & [O\,I]\,6300,6364 & OH(7,0)-(6,2) &
O$_2$(b-X)(0-1) & 0.543\,$\mu$m \\
\noalign{\smallskip}
\hline
\noalign{\smallskip}
$h_\mathrm{\,layer}$ & km &
97 & 92 & 270 & 87 & 94 & 90 \\
$N_\mathrm{spec}$ & -- &
1186 & 1046 & 1046 & 1046 & 839 & 874 \\
$\langle I \rangle$\tablefootmark{a} &
$\mathrm{phot\,s}^{-1}\,\mathrm{m}^{-2}\,\mathrm{arcsec}^{-2}$ &
3.6 & 0.81 & 3.5 & 60. & 6.0 & 80.\tablefootmark{b} \\
& &
(1.6) & (0.48) & (3.4) & (19.) & (2.5) & (31.\tablefootmark{b}) \\
& R\tablefootmark{c} &
190. & 43. & 190. & 3200. & 320. & 4.3\tablefootmark{\,d} \\
& &
(90.) & (26.) & (180.) & (1000.) & (130.) & (1.6\tablefootmark{\,d}) \\
$m_\mathrm{sun}$\tablefootmark{e} & sfu$^{-1}$ &
0.0087 & 0.0011 & 0.0068 & 0.0001 & 0.0063 & 0.0061 \\
$P_\mathrm{season,min}$\tablefootmark{f} & -- &
1 & 1 & 1 & 5 & 5 & 1 \\
$f_\mathrm{season,min}$ & -- &
0.86 & 0.54 & 0.30 & 0.82 & 0.71 & 0.81 \\
$P_\mathrm{season,max}$\tablefootmark{f} & -- &
3 & 3 & 3 & 6 & 3 & 3 \\
$f_\mathrm{season,max}$ & -- &
1.44 & 1.84 & 1.51 & 1.14 & 1.33 & 1.24 \\
\noalign{\smallskip}
\hline
\end{tabular}
\tablefoot{\\
\tablefoottext{a}{Mean intensity and variability (in parentheses) for solar
cycles 19 to 23.}\\
\tablefoottext{b}{In \radunits{}.}\\
\tablefoottext{c}{1\,R (Rayleigh) $\approx$
0.018704\,$\mathrm{phot\,s}^{-1}\,\mathrm{m}^{-2}\,\mathrm{arcsec}^{-2}$.}\\
\tablefoottext{d}{In $\mathrm{R}\,\mathrm{nm}^{-1}$.}\\
\tablefoottext{e}{Slope for solar activity correction relative to mean solar
radio flux for cycles 19 to 23 ($\approx 129$\,sfu).}\\
\tablefoottext{f}{Two-month bin with minimum or maximum correction factor:
1~=~Dec/Jan, 2~=~Feb/Mar, 3~=~Apr/May, 4~=~Jun/Jul, 5~=~Aug/Sep, 6~=~Oct/Nov.}
}
\end{table*}

The most prominent optical airglow line is [O\,I]\,5577 (see
Fig.~\ref{fig:varclasses}, panel~1). Its dominant component at an altitude of
about 97\,km is probably produced by the process described in Barth \&
Hildebrand (\cite{BAR61}). The crucial reactions are
\begin{equation}\label{eq:barth1}
\mathrm{O} + \mathrm{O} + \mathrm{M} \rightarrow \mathrm{O}^*_2 + \mathrm{M},
\end{equation}
\begin{equation}\label{eq:barth2}
\mathrm{O}^*_2 + \mathrm{O} \rightarrow \mathrm{O}_2 + \mathrm{O}(^1\mathrm{S}),
\end{equation}
and
\begin{equation}\label{eq:barth3}
\mathrm{O}(^1\mathrm{S}) \rightarrow \mathrm{O}(^1\mathrm{D}) +
557.7\,\mathrm{nm},
\end{equation}
where M is an arbitrary reaction partner. The first reaction
(Eq.~\ref{eq:barth1}) also plays an important role for the emission bands of
molecular oxygen. In order to start the three body collision, oxygen molecules
have to be split by hard UV photons:
\begin{equation}\label{eq:o2splitup}
\mathrm{O}_2 + h \nu \rightarrow \mathrm{O} + \mathrm{O}.
\end{equation}
This process mainly occurs during the day by solar UV radiation, which implies
that the oxygen airglow intensity depends on the solar activity and observing
time. The O$(^1{\rm D})$ state can lead to [O\,I]\,6300 emission (see
Fig.~\ref{fig:varclasses}, panel~3). However, in the mesopause the deactivation
of this metastable state is mostly caused by collisions. Hence, strong
[O\,I]\,6300 emission is restricted to the thermosphere at altitudes of about
270\,km. In contrast to the mesosphere, the excitation there is caused by
dissociative recombination (Bates \cite{BAT82}), i.e. it depends on the
electron density:
\begin{equation}\label{eq:disrec}
\mathrm{O}^+_2 + \mathrm{e}^- \rightarrow \mathrm{O} + \mathrm{O}(^1\mathrm{D}).
\end{equation}
The ionised molecular oxygen is produced by the charge transfer reaction
\begin{equation}\label{eq:chargetrans}
\mathrm{O}^+ + \mathrm{O}_2 \rightarrow \mathrm{O}^+_2 + \mathrm{O}.
\end{equation}
The amount of ionised oxygen and free electrons depends on the solar radiation.
The most striking features of the airglow are the Meinel OH bands (Meinel et
al. \cite{MEI54}), which dominate at red optical wavelengths and beyond (see
Fig.~\ref{fig:varclasses}, panel~4). The fundamental and overtone
rotational-vibrational transitions are mainly produced by the Bates-Nicolet
(\cite{BAT50}) mechanism
\begin{equation}\label{eq:bates}
\mathrm{O}_3 + \mathrm{H} \rightarrow \mathrm{OH}^* + \mathrm{O}_2.
\end{equation}
Most emission originates in a relatively thin layer at 87\,km. Finally, the
upper state of the prominent D lines of neutral sodium at 589.0 and 589.6\,nm
(see Fig.~\ref{fig:varclasses}, panel~2), which originate in a thin layer at
92\,km, is probably excited by
\begin{equation}\label{eq:nad1}
\mathrm{NaO} + \mathrm{O} \rightarrow \mathrm{Na}(^2\mathrm{P}) + \mathrm{O}_2
\end{equation}
(Chapman \cite{CHA39}). NaO appears to be provided by
\begin{equation}\label{eq:nad2}
\mathrm{Na} + \mathrm{O}_3 \rightarrow \mathrm{NaO} + \mathrm{O}_2.
\end{equation}

For studying airglow intensity and variability, we assigned the optical airglow
lines to five classes, for which a similiar variability is reasonable or could
be proved by a correlation analysis (see Patat \cite{PAT08}). These classes are
(1) green O\,I, (2) Na\,I\,D, (3) red O\,I, (4) OH, and (5) O$_2$ (see
Fig.~\ref{fig:varclasses}). The first three classes correspond to atomic lines,
the other two groups comprise molecular bands. The O\,I lines are divided into
two classes. The green O\,I line at 557.7\,nm mainly originates at altitudes
slightly below 100\,km, while the other significant O\,I lines (all being in
the red wavelength range) tend to originate at altitudes greater than 200\,km
(see Khomich et al. \cite{KHO08}). Moreover, the kind of chemical reactions are
also different, since the high altitude emissions are usually related to
reactions involving ions (see above). Another small group consists of the
sodium D lines and a weak K\,I line at 769.9\,nm (see
Fig.~\ref{fig:varclasses}, panel~2). By far most of the airglow lines belong to
the two classes of molecules which produce band structures by ro-vibrational
transitions. The OH electronic ground state (X-X) bands with upper vibrational
levels $v' \le 9$ cover the wavelength range longwards of about 0.5\,$\mu$m
with increasing band strength towards longer wavelengths. There are several
electronic transitions with band systems for O$_2$ in the optical.
Relatively weak bands of the Herzberg~I (A-X) and Chamberlain (A'-a) systems
originate at near-UV and blue wavelengths (see Cosby et al. \cite{COS06}). At
wavelengths longwards of 0.6\,$\mu$m, the atmospheric (b-X) band system
produces several strong bands. However, those bands related to the vibrational
$v = 0$ level at the electronic ground state X are strongly absorbed in the
lower atmosphere (see Sect.~\ref{sec:airglowext}). Therefore, the only
remaining strong band in the investigated wavelength range is O$_2$(b-X)(0-1)
at about 864.5\,nm (see Fig.~\ref{fig:varclasses}, panel~5). Variability
studies of O$_2$ have to rely on the results from this band.

For the identification of airglow lines and as an input line list for the sky
model, we use the list of Cosby et al. (\cite{COS06}). It consists of 2805
entries with information on the line wavelengths, widths, fluxes, and line
identities. It is based on the sky emission line atlas of Hanuschik
(\cite{HAN03}), which was obtained from a total of 44 high-resolution
($R \sim 45000$) VLT UVES spectra. By combining different instrumental set-ups,
the wavelength range from $0.314$ to $1.043$\,$\mu$m could be covered. The
accuracy of the wavelength calibration is better than 1\,pm (cf.
Sect.~\ref{sec:airglowext}). The UVES spectra were flux calibrated by means of
the T\"ug (\cite{TUG77}) extinction curve. Since the use of a point-source
extinction curve for the airglow leads to systematic errors (see
Sect.~\ref{sec:airglowext}) and the La Silla extinction curve was used for
Cerro Paranal, we corrected the line extinction by means of the Patat et al.
(\cite{PAT11}) extinction curve and the recipes given in
Eqs.~\ref{eq:r-airglowfit} and \ref{eq:m-airglowfit}. At the lower wavelength
limit, this caused an extreme correction factor of 0.37 and in the
range of the O$_2$(b-X)(0-0) band a maximum factor of 4.5 due to the line
molecular absorption correction (see Sect.~\ref{sec:airglowext}). However, for
the strongest 100 lines up to a wavelength of 0.92\,$\mu$m, we only obtain a
mean correction factor of 0.98. In a similar way as discussed in
Sect.~\ref{sec:data} for the Patat (\cite{PAT08}) data, we considered how
a different extinction curve affects the response curves for flux calibration.
Here, we obtain for the 100 brightest lines another correction of
0.97, which is representative of the factors for the entire wavelength range,
which range from 0.95 to 1.00. Due to a gap at about 0.86\,$\mu$m,
the Cosby et al. (\cite{COS06}) line list misses a part of the important
O$_2$(b-X)(0-1) band. There is an unpublished UVES 800U spectrum related to the
study of Hanuschik (\cite{HAN03}) that covers the gap. We used this spectrum to
measure the line intensities. Determined by a few overlapping lines, the
resulting intensities were then scaled to those in the line list. For the line
identification, we considered O$_2$ and OH line data from the HITRAN database
(see Rothman et al.~\cite{ROT09}). For the final line list for the sky model,
the intensities of each variability class were scaled to match the mean value
derived from the variability analysis (see below). Finally, the line
wavelengths were converted from air to vacuum by
\begin{equation}\label{eq:airvac}
\lambda_\mathrm{vac} = n \,\lambda_\mathrm{air}.
\end{equation}
The refractive index
\begin{equation}\label{eq:edlen}
n = 1 + 10^{-8} \left(8342.13 + \frac{2406030}{130 - \sigma^2} +
\frac{15997}{38.9 - \sigma^2}\right),
\end{equation}
where $\sigma = \lambda^{-1}$ and $\lambda$ is in $\mu$m (Edl\'en \cite{EDL66}).
This formula is also used internally in the sky model code if an output in air
wavelengths is required (see Table~\ref{tab:modelpar}).

\subsubsection{Line variability}\label{sec:linevar}

\begin{figure*}
\centering
\includegraphics[height=18.0cm,angle=-90,clip=true]{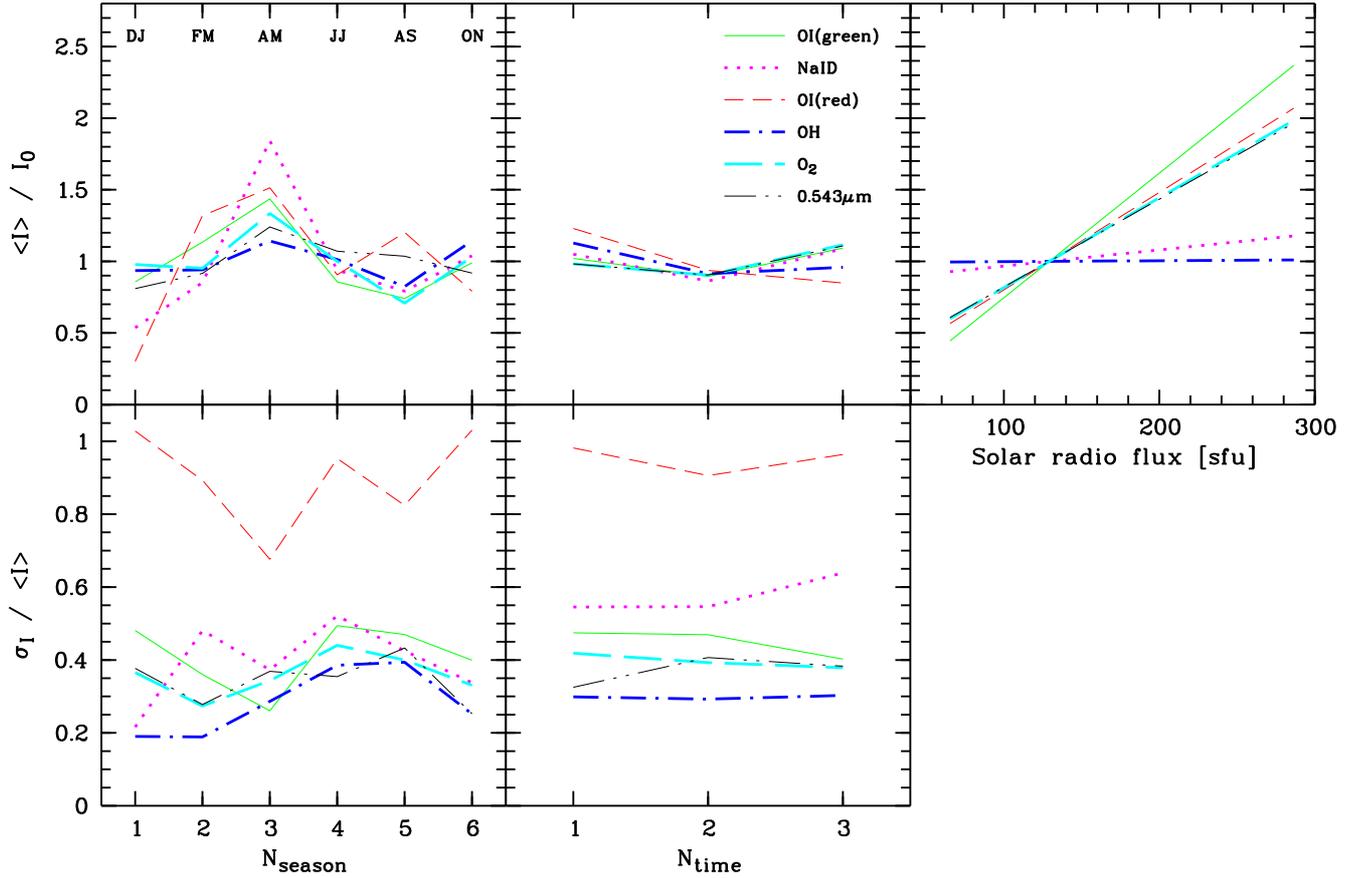}
\caption[]{Variability correction for the five airglow line classes and the
airglow continuum (analysed at 0.543\,$\mu$m) of the sky model. {\em Upper
panels:} The variability is shown as a function of the bimonthly
period (1 = Dec/Jan, ..., 6 = Oct/Nov), time bin (third of the night), and
solar activity measured by the solar radio flux. {\em Lower panels:} For
bimonthly period and time bin, the relative uncertainties of the variability
correction are displayed.}
\label{fig:illfeatvar}
\end{figure*}

\begin{figure*}
\centering
\includegraphics[height=18.0cm,angle=-90,clip=true]{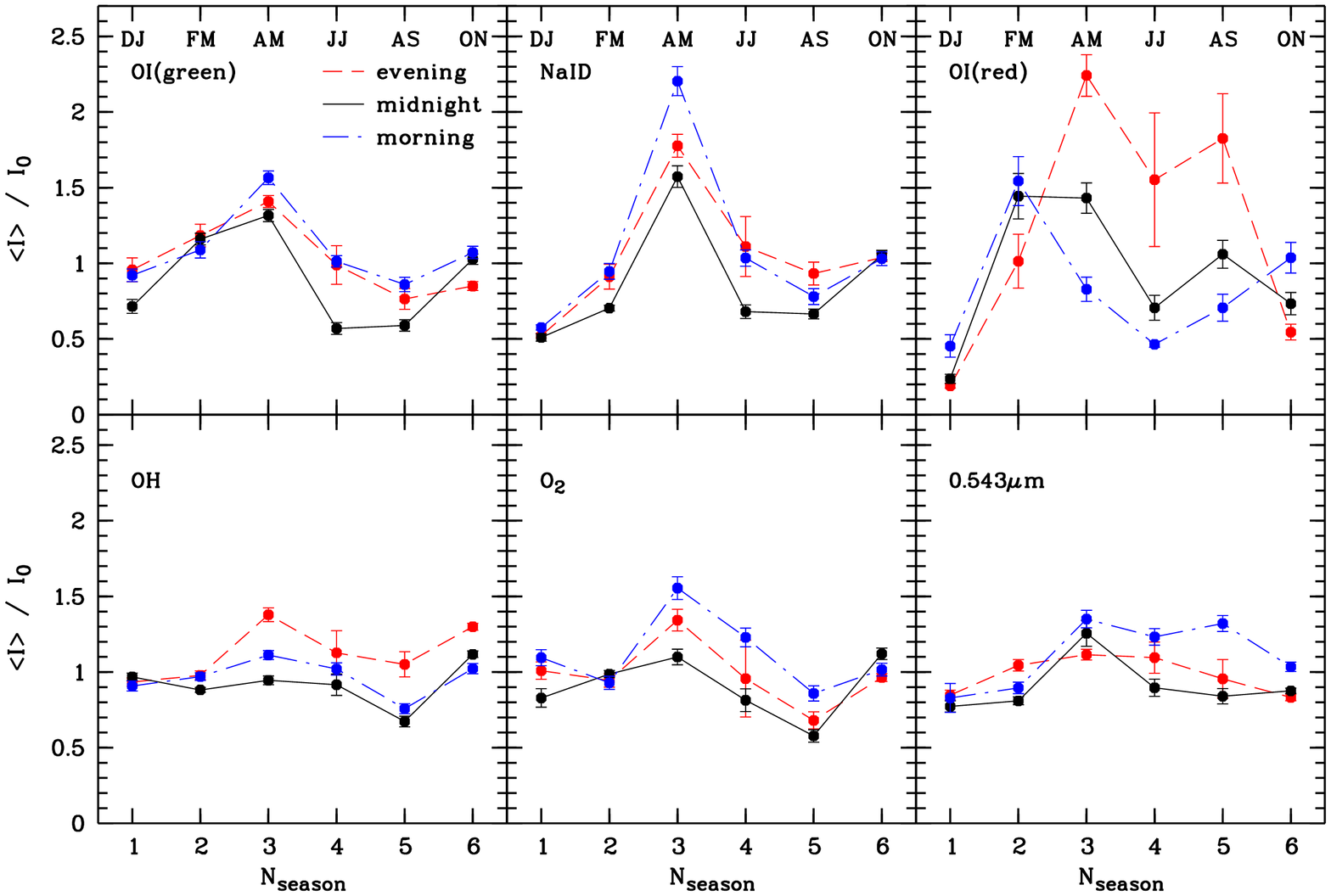}
\caption[]{Variability correction factors and their errors depending on
bimonthly period (1 = Dec/Jan, ..., 6 = Oct/Nov) and night time (third of the
night: evening, midnight, morning) for the five airglow line classes and the
airglow continuum (analysed at 0.543\,$\mu$m) of the sky model. Data points
only differing in the bimonthly period are connected by lines.}
\label{fig:illbinvar}
\end{figure*}

In general, airglow lines show strong variability on time scales ranging from
minutes to years. This behaviour can be explained by the solar activity cycle,
seasonal changes in the temperature, pressure, and chemical composition of the
emission layers, the day-night contrast, dynamical effects such as internal
gravity waves and geomagnetic disturbances (see Khomich et al. \cite{KHO08}).
In order to consider airglow variability in our sky model, we derived a
semi-empirical model based on 1186 VLT FORS\,1 sky spectra (Patat \cite{PAT08};
see Sect.~\ref{sec:data}). The demands on an airglow variability model are
twofold. First of all, it should include all major predictable variability
properties. This excludes stochastic wave phenomena like gravity waves.
Second, the derived parametrisation should be robust. For studying many
variability triggers, about 1000 spectra is not statistically a high number.
For this reason, only a few variables can be analysed. Since this neglection
leads to higher uncertainties, an ideal set of parameters has to be found that
provides statistically significant, predictable variations. For our airglow
model, we studied the effect of solar activity, period of the year, and time of
the night. The analysis was carried out for each of the five variability
classes (see Fig.~\ref{fig:varclasses}).

The first step for measuring the line and band fluxes was the subtraction of
scattered moonlight, scattered starlight, and zodiacal light by using the
recipes given in Sect.~\ref{sec:scatcomp}. The resulting spectra were then
corrected for airglow continuum extinction as described in
Sect.~\ref{sec:airglowext}. For the flux measurements, continuum windows were
defined. Their central wavelengths are listed in Table~\ref{tab:airglowcont}.
The widths were 4\,nm, except for 0.767\,$\mu$m, where 0.3\,nm was chosen to
reduce the contamination by strong OH lines and the O$_2$(b-X)(0-0) band (see
also Sects.~\ref{sec:airglowext} and ~\ref{sec:cont}). Then, the continuum
fluxes were interpolated to obtain line and band intensities. The hydroxyl
bands between 0.642 and 0.858\,$\mu$m, i.e. OH(6-1) to OH(6-2), and the
O$_2$(b-X)(0-1) band between 0.858 and 0.872\,$\mu$m were measured
automatically. In a second iteration, this procedure was improved by
subtracting the obtained O$_2$ bands from the OH bands and vice versa before
the intensity measurement. This way, the contamination by undesired lines can
be minimised. Apart from the bands, the lines [O\,I]\,5577, Na\,I\,D,
[O\,I]\,6300, and [O\,I]\,6364 were analysed. The continuum fit is very
critical for these lines, because of the narrow width of the features and the
significant contamination by OH lines. For this reason, the line measurements
were also manually performed for a subsample of 15 spectra for each
instrumental set-up. The results were then used to derive intensity-dependent
correction functions for the automatic procedure. Finally, the resulting band
and line fluxes were corrected for the discrepancy between the molecular
absorption of continuum and lines (see Sect.~\ref{sec:airglowext}). These
corrections were usually very small, since the strongest lines of each class do
not overlap with significant telluric absorption.

A variability model can be efficiently determined by deriving a reference
intensity and then applying multiplicative correction factors for each
variability parameter (cf. Khomich et al. \cite{KHO08}). As a reference
value for each variability class, we define the mean intensity at zenith for
the five solar cycles 19 to 23, i.e. the years 1954 to 2007. The zenith
intensity can be estimated by a correction for the van Rhijn effect
(Eq.~\ref{eq:vanrhijn}). The correction factors depend on the target zenith
distance and the emission layer height (see Table~\ref{tab:vardata}). For
describing the solar activity, we take the Penticton-Ottawa solar radio flux at
10.7\,cm $S_\mathrm{10.7\,cm}$ (Covington \cite{COV69}; see also
Sect.~\ref{sec:data}). For the defined period, we derive a mean flux of
129\,sfu. We can also derive a mean $S_\mathrm{10.7\,cm}$ for the spectroscopic
sample. For this purpose, we obtained the monthly $S_\mathrm{10.7\,cm}$ averages
corresponding to each spectrum. Due to the delayed response of the Earth's
atmosphere regarding solar activity of up to several weeks (see e.g. Patat
\cite{PAT08}), monthly values are more suitable than diurnal ones. In the
end, a mean $S_\mathrm{10.7\,cm}$ of 150\,sfu could be derived for the entire
FORS\,1 data set. By performing a regression analysis for the relation between
line intensity and $S_\mathrm{10.7\,cm}$ for each line class, we were able to
obtain correction factors for a scaling of the lines to intensities typical
of the reference solar radio flux. The resulting mean fluxes for the different
classes were used to correct the fluxes in the reference line list (see
Sect.~\ref{sec:linelist}). To derive the scaling factors, we integrated over
all the listed lines in a class that were measured in the spectra. This
required adding up the OH intensities of all bands from (6,1) to (6,2) and
summing up the fluxes of the [O\,I]\,6300 and [O\,I]\,6364 lines, which have a
fixed 3:1 intensity ratio. The extinction-corrected Cosby et al. (\cite{COS06})
intensities had to be corrected by factors between 0.43 for the red O\,I lines
and 1.06 for Na\,I\,D lines to reach the intensities for the reference
conditions. These reference intensities are given in Table~\ref{tab:vardata}.
They are slightly lower than those in Patat (\cite{PAT08}) due to the lower
mean solar radio flux.

The $S_\mathrm{10.7\,cm}$-dependent linear fits to the airglow intensities (see
also Patat \cite{PAT08}) are also used to parametrise a correction of solar
activity for our sky model. Even though the parametrisation is reasonable (cf.
Khomich et al. \cite{KHO08}), there may be systematic deviations for monthly
$S_\mathrm{10.7\,cm}$ beyond the data set limits of 95 and 228\,sfu (see
Sect.~\ref{sec:data}). With the dependence on solar radio flux removed,
seasonal and nocturnal variations can be investigated. Since the intensity has
a complex dependence on these parameters (see Khomich et al. \cite{KHO08}), we
divided the data set into 6 seasonal and 3 night-time bins. In the same way as
described in Sect.~\ref{sec:transmission} for the transmission curves, the
two-month periods start with the December/January combination. The night, as
limited by the astronomical twilight, is always divided into three periods of
equal length. This means that in winter the periods are longer than in summer.
For the latitude of Cerro Paranal (see Sect.~\ref{sec:introduction}), the
Dec/Jan night bins are about 30\% shorter than the Jun/Jul bins. Due to the
additional consideration of specific averages over the entire year and/or
night, the total number of bins defined for the sky model is 28 (see also
Table~\ref{tab:modelpar}). Since the number of spectra in the individual bins
differed by an order of magnitude, the average values over several bins could
be biased. Therefore, all bins were filled with the same number of spectra by
randomly adding spectra from the same bin, i.e. the data were cloned. This
approach was also used for deriving the reference intensities. As a
consequence, the mean solar radio flux of the data set changed from 153 (see
Sect.~\ref{sec:data}) to 150\,sfu. Finally, mean values and standard
deviations were calculated for the individual and multiple bins using the
original and bias-corrected data sets respectively. While the mean value
provides an intensity correction factor for seasonal and/or night-time
dependence, the standard deviation indicates the contribution of neglected
variability components to the measured intensity. It also depends on the data
quality, uncertainties in the analysis, and the accuracy of other sky model
components. Therefore, the standard deviation of each bin should be an upper
limit for the unmodelled airglow variability in the investigated data set.

The results of the airglow variability study are summarised in
Table~\ref{tab:vardata}, and Figs.~\ref{fig:illfeatvar} and
\ref{fig:illbinvar}. The solar activity strongly influences the intensities of
O\,I and O$_2$, where the slopes are 0.0063 to 0.0087\,sfu$^{-1}$. On the other
hand, weak or no significant correlations are present for sodium and hydroxyl.
These findings are in qualitative agreement with the relations given in Khomich
et al. (\cite{KHO08}), who provide 0.0025 to 0.0060\,sfu$^{-1}$ for the
first and about 0.0015\,sfu$^{-1}$ for the second group from measurements at
different observing sites (mainly in the Northern Hemisphere). The relatively
low value of 0.0025\,sfu$^{-1}$ for the green O\,I line compared to
0.0087\,sfu$^{-1}$ from our model is not critical. A more precise
latitude-dependent analysis of satellite-based data by Liu \& Shepherd
(\cite{LIU08}) resulted in 0.0065\,sfu$^{-1}$ for the latitude range from
$-20\degr$ to $-30\degr$. For $-30\degr$ to $-40\degr$ they even obtain the
same slope as in our model. The discrepancies in the results from literature
show that quantitative comparisons are difficult with different observing
sites, observing periods, instruments, and analysis methods. Solar activity
directly affects the Earth's atmosphere by the amount of hard UV photons.
The higher the photon energy, the more the production rate depends on solar
activity (see Nicolet \cite{NIC89}). This relation appears to be important for
O\,I and O$_2$ airglow emission, which depends on the dissociation or
ionisation of molecular oxygen by hard UV photons (see
Sect.~\ref{sec:linelist}).

Significant dependence on the bimonthly period is found for all line classes
(see Fig.~\ref{fig:illfeatvar}). Na\,I\,D and the red O\,I lines indicate the
strongest variability with a maximum intensity ratio of 3.4 and 5.0
respectively, whereas OH shows the weakest dependence with a ratio of 1.4 (see
Table~\ref{tab:vardata}). Except for OH, the maximum is reached in the autumn
months April and May. For the atomic lines, the minimum arises in Dec/Jan, and
for the molecular bands, it appears to be in Aug/Sep. The variations found
reflect the semi-annual oscilliations of airglow intensity (cf. Patat
\cite{PAT08}), which are characterised by two maxima and minima of different
strength per year. Complex empirical relations based on long-term observations
at many observing sites as given by Khomich et al. (\cite{KHO08}) can be used
to estimate this intensity variation. However, since these relations depend on
rough interpolations and extrapolations, they can only provide a coarse
variability pattern for an arbitrary location on Earth. Nevertheless, these
relations confirm the amplitude of the variations and the rough positions of
the maxima and minima. For example, the peak-to-peak ratios for green O\,I,
Na\,I\,D, and OH are 1.4, 3.8, and 1.3, which agrees well with the sky model
values 1.7, 3.4, and 1.4 (see Table~\ref{tab:vardata}). Apart from the obvious
change of the incident solar radiation, the seasonal variations are related to
changes in the layer heights of the atmospheric constituents, changes in the
temperature profiles, and the dynamics of the upper atmosphere in general.

Averaged over the year, the dynamical range of the night-time variations is in
the order of 20 to 30\%, which is distinctly smaller than the seasonal
variations. Only the red O\,I lines show a clear decrease by a factor of 1.4
from the beginning to the end of the night. Usually, the diurnal variations
are the largest in airglow intensity with possible amplitudes of several orders
of magnitude (see Khomich et al. \cite{KHO08}). Since we exclude daytime and
twilight conditions with direct solar radiation, the variability is much
smaller. In addition, the use of only three bins can smooth out the intensity
variations. However, an investigation with the unbinned data (see also Patat
\cite{PAT08}) and results from photometric studies (Mattila et al.
\cite{MAT96}; Benn \& Ellison \cite{BEN98}; Patat \cite{PAT03}) confirm the
absence of strong trends. Averaging over all bimonthly periods appears to
cancel part of the variability. This is illustrated in
Fig.~\ref{fig:illbinvar}, which shows the bimonthly dependence of the airglow
intensity for each third of the night. At midnight, i.e. the maximum time
distance to the twilight, the intensities of the mesopause lines tend to reach
a minimum. However, the deviations from the intensities of the evening and
morning bins appear to be a function of season. While in winter the differences
reach a maximum, summer has the smallest differences. This behaviour can be
explained by the night lengths at Cerro Paranal, which can vary by more than
30\%. In summer, the average length of time from the closest twilight is
significantly smaller than in winter. This dependence supports the common
treatment of seasonal and night-time variations in the sky model.

The residual statistical variations of the airglow model, as depicted in
Fig.~\ref{fig:illfeatvar}, are mainly between 20 and 50\% for the lines
originating in the mesopause region. The smallest uncertainties are found for
OH in summer, whereas the ionospheric O\,I lines indicate very strong
unpredictable variations with an amplitude comparable to the intensity itself
(see also Patat \cite{PAT08}). This behaviour is caused by the aurora-like
intensity fluctuations from geomagnetic disturbances. This effect is
particularly important for Cerro Paranal, because of its location close to one
of the two active regions about $20\degr$ on either side of the geomagnetic
equator (Roach \& Gordon \cite{ROA73}).

\subsection{Airglow continuum}\label{sec:cont}

The airglow continuum is the least understood emission component of the night
sky. In the optical, the best-documented process is a chemiluminescent reaction
of nitric oxide and atomic oxygen in the mesopause region as proposed by
Krassovsky (\cite{KRA51}):
\begin{equation}\label{eq:noo}
\mathrm{NO} + \mathrm{O} \rightarrow \mathrm{NO}_2 + h \nu
\end{equation}
\begin{equation}\label{eq:no2}
\mathrm{NO}_2 + \mathrm{O} \rightarrow \mathrm{NO} + \mathrm{O}_2.
\end{equation}
This reaction appears to dominate the visual and is expected to have a broad
maximum at about 0.6\,$\mu$m (see Sternberg \& Ingham \cite{STE72a}; von
Savigny et al. \cite{SAV99}; Khomich et al. \cite{KHO08}). Reactions of NO and
ozone could be important for the airglow continuum at red to near-IR
wavelengths (Clough \& Thrush \cite{CLO67}; Kenner \& Ogryzlo \cite{KEN84}).
In particular, the reaction
\begin{equation}\label{eq:noo3}
\mathrm{NO} + \mathrm{O}^*_3 \rightarrow \mathrm{NO}^*_2 + \mathrm{O}_2
\end{equation}
is expected to significantly contribute to the optical by continuum emission
with a broad maximum at about 0.85\,$\mu$m (Kenner \& Ogryzlo \cite{KEN84};
Khomich et al. \cite{KHO08}). Finally, a pseudo continuum by molecular band
emission produced by
\begin{equation}\label{eq:feo3}
\mathrm{Fe} + \mathrm{O}_3 \rightarrow \mathrm{FeO}^* + \mathrm{O}_2
\end{equation}
(West \& Broida \cite{WES75}) could be an important component of the continuum
between 0.55 and 0.65\,$\mu$m (Jenniskens et al. \cite{JEN00}; Evans et al.
\cite{EVA10}; Saran et al. \cite{SAR11}).

\begin{table}
\caption[]{Airglow/residual continuum and its intensity variation relative to
the reference wavelength 0.543\,$\mu$m.}
\label{tab:airglowcont}
\centering
\vspace{5pt}
\begin{tabular}{c c c c}
\hline\hline
\noalign{\smallskip}
$\lambda$ [$\mu$m] & $\langle I \rangle / I_0$ & $\sigma_I / I_0$ &
$\sigma_I / \langle I \rangle$ \\
\noalign{\smallskip}
\hline
\noalign{\smallskip}
0.369 & 0.92 & 0.13 & 0.14 \\
0.387 & 0.65 & 0.11 & 0.17 \\
0.420 & 0.69 & 0.10 & 0.15 \\
0.450 & 0.70 & 0.11 & 0.15 \\
0.480 & 0.79 & 0.11 & 0.14 \\
0.510 & 0.79 & 0.09 & 0.12 \\
0.543 & 1.00 & 0.00 & 0.00 \\
0.575 & 1.41 & 0.25 & 0.18 \\
0.608 & 1.40 & 0.29 & 0.20 \\
0.642 & 1.18 & 0.19 & 0.16 \\
0.675 & 1.14 & 0.13 & 0.12 \\
0.720 & 1.32 & 0.21 & 0.16 \\
0.767 & 1.64 & 0.46 & 0.28 \\
0.820 & 2.57 & 0.67 & 0.26 \\
0.858 & 3.56 & 1.06 & 0.30 \\
0.872 & 3.76 & 1.12 & 0.30 \\
\noalign{\smallskip}
\hline
\end{tabular}
\tablefoot{ The reference continuum intensity at 0.543\,$\mu$m is
79.8\,\radunits{} or 4.27\,$\mathrm{R}\,\mathrm{nm}^{-1}$.}
\end{table}

\begin{figure}
\centering
\includegraphics[width=8.8cm,clip=true]{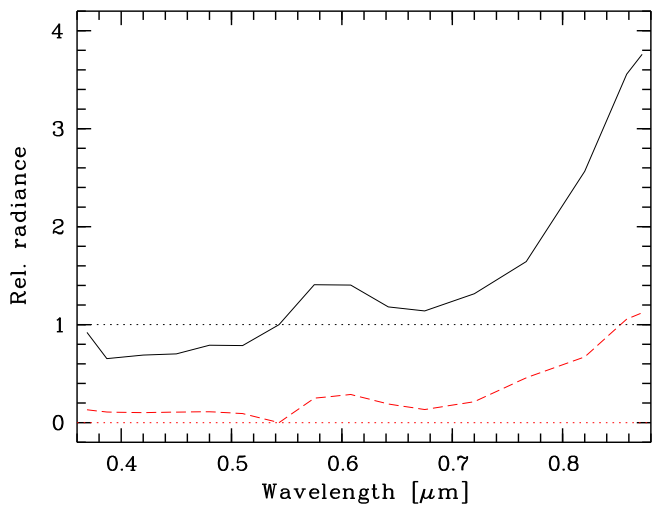}
\caption[]{Airglow/residual continuum (solid line) and its variability (dashed
line) relative to the reference wavelength $0.543$\,$\mu$m.}
\label{fig:airglowcont}
\end{figure}

The airglow continuum was analysed in a similar way as the airglow emission
lines. As reported in Sect.~\ref{sec:linevar}, the derivation of line
intensities already required the measurement of airglow continuum fluxes in
suitable windows (see Table~\ref{tab:airglowcont}). These results were refined
by subtracting possible contributions from airglow lines by applying the line
emission model as described in Sect.~\ref{sec:lines}. For the variability
analysis, the Patat (\cite{PAT08}) data set (see Sect.~\ref{sec:data}) was
restricted to the 874 FORS\,1 spectra, where the Moon was below the horizon.
This avoids large errors in the continuum flux from the subtraction of a bright
and uncertain component (see Sect.~\ref{sec:moon}). Even though the absolute
uncertainties of the other components are significantly smaller, the derived
continuum is a residual continuum. It is affected by uncertainties in zodical
light brightness, scattered starlight, airglow line emission, molecular
absorption, possible unconsidered minor components, scattered light from
dispersive elements\footnote{This effect can cause line profiles with wide
Lorentzian wings (see Ellis \& Bland-Hawthorn \cite{ELL08}). However, we did
not detect significant wings in the profiles of the strong airglow lines or
lines in the wavelength calibration lamp spectra. In view of the relatively low
contrast between lines and continuum for our spectra (cf. the near-IR regime),
we can conclude that a possible continuum contamination of more than a few per
cent is unlikely, even at the reddest wavelengths.}, and flux calibration of
the spectra. While the other sky model components should cause an uncertainty
in the order of a few per cent only (for 0.767\,$\mu$m this could be somewhat
higher, see Sect.~\ref{sec:linevar}), the uncertainty in the flux calibration
could be significantly more. Since theoretical components are subtracted from
the observed spectra to obtain the airglow continuum, errors in the flux
calibration have a larger effect on the resulting continuum than on the total
spectra. For example, for a 50\% contribution of the continuum to the total
flux, a flux error of 10\% would convert into a continuum error of 20\%.

For the subsequent analysis, we defined a reference continuum wavelength. Due
to its availability in all spectroscopic modes and the absence of contaminating
emission lines, we selected 0.543\,$\mu$m. In the following, we studied the
continuum relative to this wavelength, i.e. the mean continuum was derived
by scaling the fluxes to the reference solar radio flux of 129\,sfu (see
Sect.~\ref{sec:linevar}) by using only the linear regression results and
binning corrections derived for 0.543\,$\mu$m. This approach means that a fixed
shape of the continuum is assumed. This simplified variability model can be
more easily used in the sky model. As shown by the last column of
Table~\ref{tab:airglowcont}, there are relatively low variations of the
continuum flux relative to the reference flux for the investigated data set.
For most wavelengths, the standard deviation divided by the mean flux
$\sigma_I / \langle I \rangle$ is lower than 20\% and only reaches 30\% at the
reddest wavelengths. In view of all the uncertainties discussed above, which
also contribute to these numbers, the assumption of a fixed continuum slope is
sufficient for an application for the ETC sky background model. This assumption
is also supported by the high correlation factors for different continuum
windows measured by Patat (\cite{PAT08}). It should be noted that the number of
available spectra decreases towards the margins in the analysed spectral range
due to the different instrumental set-ups (see Sect.~\ref{sec:data}). In
particular, wavelengths shortwards of 0.44\,$\mu$m have only 239 spectra, i.e.
27\% of the sample used for 0.543\,$\mu$m. The change in the sample size as
well as different systematic errors in the response curves for the different
set-ups cause systematic uncertainties in the airglow continuum shape. By
comparing the mean fluxes in overlapping wavelength regions of the different
set-ups, we estimate an uncertainty in the order of 10\%.

We obtained a mean airglow continuum for Cerro Paranal with an intensity of
about 4.3\,$\mathrm{R}\,\mathrm{nm}^{-1}$ at 0.543\,$\mu$m (for R[ayleigh]
units see Table~\ref{tab:vardata}). The corresponding variability is
1.6\,$\mathrm{R}\,\mathrm{nm}^{-1}$. Considering the latitude of Cerro Paranal,
this result is in good agreement with observations that suggest an increase of
the airglow continuum from $1 - 2\,\mathrm{R}\,\mathrm{nm}^{-1}$ at the equator
to $10 - 20\,\mathrm{R}\,\mathrm{nm}^{-1}$ in the polar region for similar
wavelengths (see Davis \& Smith \cite{DAV65}; Khomich et al. \cite{KHO08}). As
illustrated in Fig.~\ref{fig:airglowcont} and listed in
Table~\ref{tab:airglowcont}, the mean airglow continuum is characterised by
a broad bump at about 0.6\,$\mu$m and an increase in intensity towards longer
wavelengths with a steep slope at about 0.8\,$\mu$m. The continuum in the bump
region appears to be mainly produced by two processes: a relatively smooth
$\mathrm{NO} + \mathrm{O}$ continuum (see Eq.~\ref{eq:noo}) and a more
structured $\mathrm{Fe} + \mathrm{O}_3$ continuum (see Eq.~\ref{eq:feo3})
showing a peak at about 0.6\,$\mu$m (see Evans et al. \cite{EVA10}). If the
bump, which indicates a width of about 0.1\,$\mu$m, could be completely
explained by the iron oxide reaction, this component would contribute up to
$1/3$ of the mean airglow continuum flux (see also Sect.~\ref{sec:spec}).
Towards longer wavelengths, it is not clear whether the observed increase in
intensity is mainly caused by relatively narrow peaks or a smooth increase (see
Sternberg \& Ingham \cite{STE72a}; Sobolev \cite{SOB78}; Content \cite{CON96}).
Up to 0.872\,$\mu$m, our measurements suggest a monotonic increase beyond the
0.6\,$\mu$m bump. Intensities of about 15\,$\mathrm{R}\,\mathrm{nm}^{-1}$, as
found by Sternberg \& Ingham (\cite{STE72a}) and Sobolev (\cite{SOB78}) at
wavelengths longwards of $0.8$\,$\mu$m, are in good agreement with our mean
spectrum, with the same intensity at about $0.86$\,$\mu$m. Also, the
intensities of Sternberg \& Ingham (\cite{STE72a}) and of the sky model mean
airglow continuum are very similar at our reference wavelength in the $V$ band.
A possible production meachanism for the strong increase at about 0.8\,$\mu$m
could be a reaction of nitric oxide with excited ozone (see Eq.~\ref{eq:noo3}),
for which a peak wavelength of about 0.85\,$\mu$m is expected (see Kenner \&
Ogryzlo \cite{KEN84}; Khomich et al. \cite{KHO08}). Then, the strong intensity
increase would be the blue wing of the bump from this reaction. The slope does
tend to decrease where the peak is predicted, and the continuum variations
there relative to 0.543\,$\mu$m (see $\sigma_I / \langle I \rangle$ in
Table~\ref{tab:airglowcont}) are the highest. This could be another argument
for different chemical processes producing the emissions at 0.55 and
0.85\,$\mu$m.

As in the case of the airglow lines, the intensity of the sky model airglow
continuum measured at $0.543$\,$\mu$m depends on zenith distance, solar radio
flux, period of the year, and period of the night. The results are shown in
Table~\ref{tab:vardata}, and Figs.~\ref{fig:illfeatvar} and \ref{fig:illbinvar}.
The dependence of the continuum emission on solar radio flux is significant and
comparable to O\,I (green and red) and O$_2$. Since OH does not indicate a
significant dependence on $S_\mathrm{10.7\,cm}$, the continuum must be the main
source for the dependence of the broad-band fluxes on solar activity (cf.
Walker \cite{WAL88}; Patat \cite{PAT03}, \cite{PAT08}). Concerning the seasonal
and nocturnal variations, the continuum at $0.543$\,$\mu$m tends to show the
best agreement with OH, i.e. the variability of these parameters is relatively
low. The maximum ratio of the bimonthly mean intensities is only about 1.5 (cf.
Patat~\cite{PAT08}).

\section{Discussion}\label{sec:discussion}

After describing all the optical model components in the previous sections,
we now discuss the quality of the sky model as a whole by comparing it to the
Patat (\cite{PAT08}) data set of sky spectra and results found in the
literature. First, sky model spectra and observed spectra are compared
(Sect.~\ref{sec:spec}). Second, a systematic quality evaluation by means of
broad-band photometry of our model with data, and a comparison to the night-sky
brightnesses from the literature are discussed in Sect.~\ref{sec:photo}.
Finally, we evaluate the improvements achieved by our sky model compared to the
current ESO ETC sky background model in the optical (Sect.~\ref{sec:etc}).

\subsection{Comparison of spectra}\label{sec:spec}

\begin{figure}
\centering
\includegraphics[width=8.8cm,clip=true]{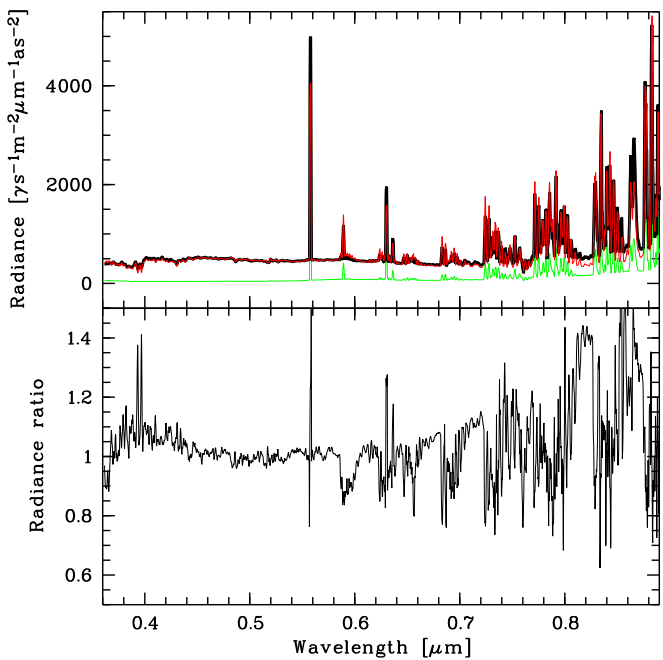}
\caption[]{Comparison of the sky model ({\em upper panel:} thick black line)
and an observed FORS\,1 300V spectrum (thin red line) with moderate lunar
contribution (see Table~\ref{tab:modelpar}). The uncertainty of the sky model
due to airglow variability is also shown (green). The {\em lower panel}
exhibits the ratio of the sky model and the observed spectrum.}
\label{fig:fors_0135}
\end{figure}

\begin{figure}
\centering
\includegraphics[width=8.8cm,clip=true]{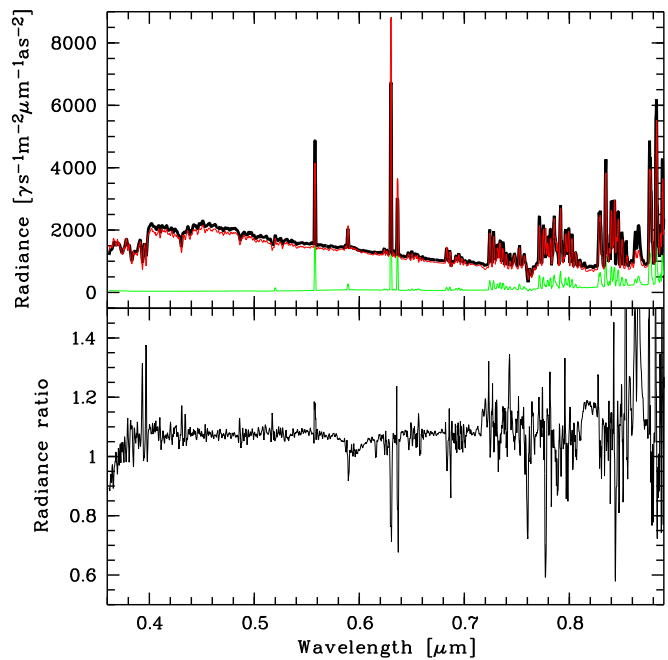}
\caption[]{Comparison of the sky model and an observed FORS\,1 300V spectrum
with strong lunar contribution. For an explanation of the different curves,
see Fig.~\ref{fig:fors_0135}.}
\label{fig:fors_0598}
\end{figure}

\begin{figure}
\centering
\includegraphics[width=8.8cm,clip=true]{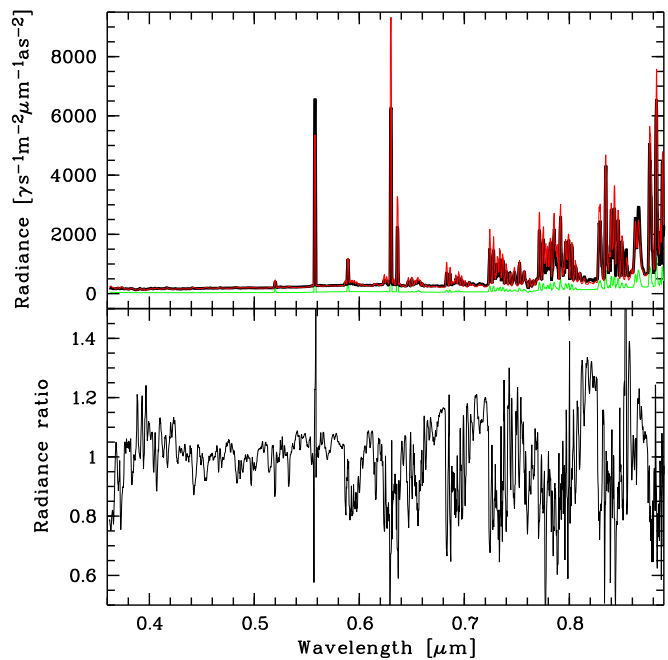}
\caption[]{Comparison of the sky model and an observed FORS\,1 300V spectrum
without lunar contribution, but strong airglow emission lines. For an
explanation of the different curves, see Fig.~\ref{fig:fors_0135}.}
\label{fig:fors_0582}
\end{figure}

\begin{figure}
\centering
\includegraphics[width=8.8cm,clip=true]{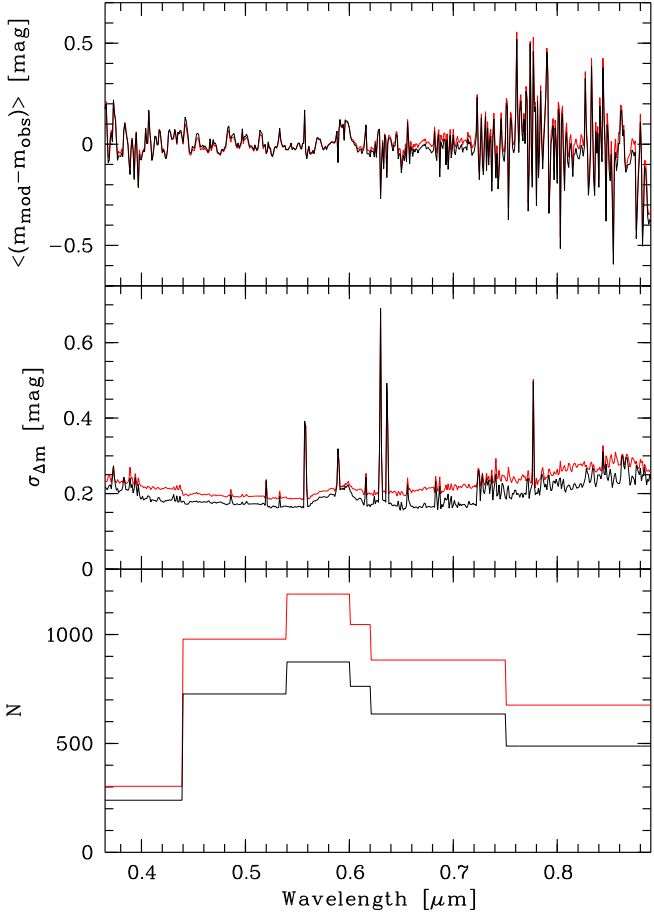}
\caption[]{Deviations between the sky model and the observed FORS\,1 spectra in
magnitudes. The mean magnitude difference ({\em upper panel}), the standard
deviation ({\em middle panel}), and the wavelength-dependent number of
considered spectra ({\em lower panel}) are shown for 1\,nm bins. Results for
the full spectroscopic data set (red) and for spectra with the Moon below the
horizon (black) are displayed.}
\label{fig:compfullspec}
\end{figure}

In order to evaluate the quality of our sky model, spectra were computed for
the different instrumental set-ups and observing conditions of the Patat
(\cite{PAT08}) data set (see Sect.~\ref{sec:data}). A detailed description of
the sky model input parameters can be found in Table~\ref{tab:modelpar}. Before
the full data set is discussed, we show a few typical sky model spectra in
comparison with the corresponding observed spectra (Figs.~\ref{fig:fors_0135}
to \ref{fig:fors_0582}).

In the optical, the relevant sky model components are scattered moonlight,
scattered starlight, zodiacal light, and airglow lines and continuum of the
upper atmosphere (see Fig.~\ref{fig:logcomp}). Figure~\ref{fig:fors_0135}
compares a typical FORS\,1 spectrum exhibiting moderate Moon contribution to
the corresponding sky model. This and the FORS\,1 spectra of the other figures
were taken with the 300V grism without order separation filter. This set-up
provides the maximum wavelength range. Due to the missing order separation
filter, there can be a slight ($\la 10\%$) contamination by the second order
spectrum at red wavelengths. However, this does not affect the quality of the
sky model (see Sect.~\ref{sec:photo}), because of the small number of spectra
used in the analysis. Figure~\ref{fig:fors_0135} demonstrates that the
different sky model components agree reasonably well with the observed
spectrum. Apart from the reddest part of the spectrum, deviations of 10 to 20\%
are typical. The best agreement is found for the continuum in the visual.
Residuals of strong sky lines in the ratio plot in the lower panel appear more
prominent than the real values due to slight differences in the instrumental
profiles of the model and the observed spectrum.

Figure~\ref{fig:fors_0598} shows the effect of a strong Moon contribution on
the sky model. Scattered moonlight dominates the continuum at all wavelengths
in the displayed range. At blue wavelengths, the example spectrum reaches up to
2000\,\radunits{}. In contrast, during moonless nights the values are about one
order of magnitude lower (see Fig.~\ref{fig:fors_0582}). Even though the
deviations between the sky model and the observed spectrum are not larger than
10\% for most of the continuum, there is a systematic trend. For a wide
wavelength range, the sky model appears to be too bright by a relatively
constant factor. Apart from deviations of the Moon illuminance from the model
in Eq.~\ref{eq:moonillu}, this discrepancy could be explained by uncertainties
in the Mie scattering model, since this changes slowly with wavelength (see
Fig.~\ref{fig:transcurve}). Moreover, the angular distance between Moon and
target $\rho = 44\degr$ is small enough for significant Mie scattering.
Similar trends with positive as well as negative deviations are observed in
other cases of high lunar sky brightness. As already mentioned in
Sect.~\ref{sec:moon}, the amplitude of the differences tends to increase for
decreasing $\rho$. Nevertheless, the uncertainties of the extended Krisciunas
\& Schaefer (\cite{KRI91}) model are usually not critical for spectra with low
to moderate lunar contribution, as it is illustrated in
Fig.~\ref{fig:fors_0135}.

A spectrum without scattered moonlight is shown in Fig.~\ref{fig:fors_0582}. In
this case, the contiuum is mainly a combination of zodiacal light and airglow
continuum. Scattered starlight only adds a few per cent to the total continuum
emission. In the $B$ and $V$ bands, the continuum of dark-time exposures is
usually well reproduced by the sky model. At longer wavelengths, the
uncertainties increase from the significant (and partly unpredictable)
variability of the airglow continuum and the simplifying assumption of a fixed
continuum shape (see Sect.~\ref{sec:cont}). The three spectra shown indicate
deviations between 20 to 40\% at about 0.82\,$\mu$m. By chance, the sky model
is brighter than the observed spectrum in all cases. Airglow emission lines can
vary significantly, depending on the line class, observing direction, solar
activity, season, and time of night. Even though there is always a stochastic
component, which cannot be modelled, the sky model accounts for such variations
quite well. The uncertainty of the line modelling for the spectra shown is in
the order of 20\%. The spectra in Figs.~\ref{fig:fors_0135} and
\ref{fig:fors_0582} were taken at the same solar radio flux (205\,sfu) and in
the last third of the night, i.e. the conspicuous differences between both
spectra can mainly be explained by seasonal variations. This is confirmed by
the fact that the sky model can roughly follow the striking changes in the
intensities of the red O\,I lines (cf. FM and JJ periods for red O\,I morning
intensities in Fig.~\ref{fig:illbinvar}).

For a more systematic comparison between the model and the data, we computed
the mean and standard deviations for the entire sample and a subset with the
Moon above the horizon. The number of spectra involved depends on the
wavelength and does not exceed 1186 and 874 respectively. The results are
displayed in Fig.~\ref{fig:compfullspec} in magnitudes. For both samples, the
systematic model deviations (upper panel) are close to zero. This is very
satisfying, but not unexpected. The airglow line and continuum components were
derived based on the FORS\,1 spectra, which are also used in the comparison.
The lack of clear spectral features in the upper panel of
Fig.~\ref{fig:compfullspec} suggests that the model quality is similar at all
wavelengths. As already discussed in Sect.~\ref{sec:cont}, the real accuracy of
the model also depends on the quality of the flux calibration of the Patat
(\cite{PAT08}) data. The uncertainties can be in the order of 10\%, and so this
appears to be the main source of systematic errors in the model intensities.

The standard deviation of the magnitude differences between the model and
observed data $\sigma_{\Delta m}$, shown in the middle panel of
Fig.~\ref{fig:compfullspec}, indicates an interesting wavelength dependence.
For the dark-time sample, the continuum uncertainty ranges between 0.16 and
0.24\,mag. Including observations with scattered moonlight increases these
values to 0.19 and 0.28\,mag. The additional variation is mainly caused by a
few spectra with large uncertainties in the Mie scattering of moonlight (see
above). These results suggest that the relative strong continuum deviations at
0.82\,$\mu$m, as in Figs.~\ref{fig:fors_0135} and \ref{fig:fors_0582}, are
relatively high compared to the entire sample. The spectrum of
$\sigma_{\Delta m}$ shows a relatively broad feature at about 0.6\,$\mu$m, which
is reminiscent of the shape of the FeO pseudo continuum (see Jennikens et al.
\cite{JEN00}; Evans et al. \cite{EVA10}; Saran et al. \cite{SAR11}). This
result supports the assumption that the airglow continuum bump at 0.6\,$\mu$m
(see Fig.~\ref{fig:airglowcont}) is mainly produced by iron oxide (see
Sect.~\ref{sec:cont}) and varies differently than the underlying NO$_2$
continuum, for which the intensitiy changes at 0.543\,$\mu$m are considered by
the sky model. Fig.~\ref{fig:compfullspec} indicates a FeO-related variability
of about 15\%. Since the mean bump contributes about 20\% of the total flux at
0.6\,$\mu$m, the variation is almost comparable to it. These strong variations
are in agreement with previous results (see Evans et al. \cite{EVA10}; Saran et
al. \cite{SAR11}). A minor source of continuum variability at 0.6\,$\mu$m is
the Chappuis ozone bands (see Fig.~\ref{fig:transcurve}), which indicate a
variation of 0.02\,mag~airmass$^{-1}$ (see Patat et al. \cite{PAT11}). The
middle panel of Fig.~\ref{fig:compfullspec} does not exhibit strong signatures
of OH bands (see Fig.~\ref{fig:varclasses}, panel~4). The bands are clearly
detectable only in the range of low continuum uncertainty. Consequently, the
quality of the OH modelling is comparable or even better than the continuum
modelling. This is particularly important for the quality of the full sky model
at long wavelengths, where OH emission dominates the spectra. The sky model is
less able to model atomic lines, as shown in Fig.~\ref{fig:compfullspec}. The
maximum variability of 0.7\,mag is found at [O\,I]\,6300. In addition, other
lines of the thermospheric ionosphere, which are usually difficult to recognise
(see Fig.~\ref{fig:varclasses}, panel~3), significantly contribute to
$\sigma_{\Delta m}$. For example, N\,I at 520\,nm and O\,I at 616, 656, and
777\,nm are clearly detectable. This confirms the conclusion of
Sect.~\ref{sec:linevar}, stating that the intensities of ionospheric lines are
difficult to predict due to their sensitivity to geomagnetic disturbances (see
Roach \& Gordon \cite{ROA73}).

\subsection{Comparison of photometry}\label{sec:photo}

\begin{figure}
\centering
\includegraphics[width=8.8cm,clip=true]{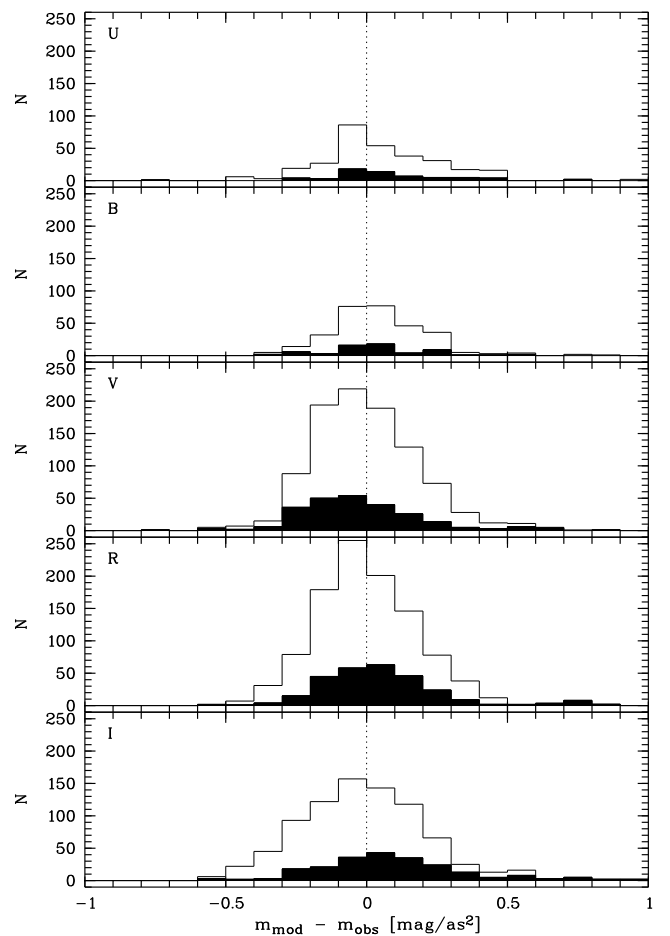}
\caption[]{Histograms for the deviations of the sky model from the FORS\,1 data
in mag for the filters $U$, $B$, $V$, $R$, and $I$. The filled histograms show
only data with the Moon above the horizon.}
\label{fig:magdiff}
\end{figure}

\begin{table*}
\caption[]{Comparison of the sky model and the Walker (\cite{WAL87})
magnitudes with the Patat (\cite{PAT08}) FORS\,1 data.}
\label{tab:magdiff}
\centering
\vspace{5pt}
\begin{tabular}{c @{\qquad} c c c c c c @{\qquad} c c c c c c}
\hline\hline
\noalign{\smallskip}
Filter & \multicolumn{6}{c}{All} & \multicolumn{6}{c}{No Moon} \\
& $N$ & $\langle \Delta m \rangle$ & $\sigma_{\Delta m}$ & $\sigma_{\rm obs}$ &
$\langle \Delta m_\mathrm{w} \rangle$ & $\sigma_{\Delta m, \mathrm{w}}$
& $N$ & $\langle \Delta m \rangle$ & $\sigma_{\Delta m}$ & $\sigma_{\rm obs}$ &
$\langle \Delta m_\mathrm{w} \rangle$ & $\sigma_{\Delta m,\mathrm{w}}$ \\
\noalign{\smallskip}
\hline
\noalign{\smallskip}
$U$ &  303 & $+0.053$ & 0.230 & 0.521 & $-0.113$ & 0.725 &
       239 & $+0.042$ & 0.213 & 0.214 & $-0.059$ & 0.703 \\
$B$ &  303 & $+0.046$ & 0.205 & 0.561 & $+0.269$ & 0.430 &
       239 & $+0.048$ & 0.174 & 0.182 & $+0.196$ & 0.407 \\
$V$ &  979 & $+0.000$ & 0.192 & 0.563 & $+0.356$ & 0.409 &
       727 & $+0.006$ & 0.175 & 0.255 & $+0.244$ & 0.322 \\
$R$ & 1046 & $+0.009$ & 0.199 & 0.447 & $+0.169$ & 0.365 &
       762 & $-0.009$ & 0.177 & 0.260 & $+0.083$ & 0.287 \\
$I$ &  839 & $-0.003$ & 0.238 & 0.365 & $+0.413$ & 0.326 &
       615 & $-0.040$ & 0.208 & 0.287 & $+0.365$ & 0.294 \\
\noalign{\smallskip}
\hline
\end{tabular}
\tablefoot{ Mean values $\langle \Delta m \rangle$ and standard deviations
$\sigma_{\Delta m}$ for magnitude differences between the model and observed data
($\Delta m = m_\mathrm{model} - m_\mathrm{obs}$) and standard deviations of the
magnitudes of the observed data $\sigma_{\rm obs}$ are shown. Moreover, mean
values $\langle \Delta m_\mathrm{w} \rangle$ and standard deviations
$\sigma_{\Delta m, \mathrm{w}}$ for magnitude differences between the Moon phase
related data of Walker and the Patat data are displayed. Results are listed for
the full and dark-time samples. The number of spectra involved is indicated by
$N$.}
\end{table*}

\begin{table*}
\caption[]{Typical sky model and literature night-sky brightnesses in
mag~arcsec$^{-2}$ for zenith, no Moon, faint zodiacal light, and different
solar radio fluxes.}
\label{tab:refmag}
\centering
\vspace{5pt}
\begin{tabular}{l l c c c c c c}
\hline\hline
\noalign{\smallskip}
Source & Site & $S_\mathrm{10.7\,cm}$\tablefootmark{a} &
$U$ & $B$ & $V$ & $R$ & $I$ \\
\noalign{\smallskip}
\hline
\noalign{\smallskip}
Benn \& Ellison (\cite{BEN98})   & La~Palma      & 80    & 22.0 & 22.7 & 21.9 &
21.0 & 20.0 \\
Walker (\cite{WAL87})            & Cerro Tololo  & 90    & 22.0 & 22.7 & 21.8 &
20.9 & 19.9 \\
Krisciunas et al. (\cite{KRI07}) & Cerro Tololo  & 130   & 22.1 & 22.8 & 21.8 &
21.2 & 19.9 \\
Mattila et al. (\cite{MAT96})    & La~Silla      & 150   &      & 22.8 & 21.7 &
20.8 & 19.5 \\
Patat (\cite{PAT08})             & Cerro Paranal & 160   & 22.4 & 22.7 & 21.7 &
20.9 & 19.7 \\
Patat (\cite{PAT03})             & Cerro Paranal & 180   & 22.3 & 22.6 & 21.6 &
20.9 & 19.7 \\
\noalign{\smallskip}
\hline
\noalign{\smallskip}
Sky model\tablefootmark{\,b}     & Cerro Paranal & 90    & 22.3 & 22.9 & 22.0 &
21.2 & 19.8 \\
                                 &               & 130   & 22.1 & 22.8 & 21.8 &
21.0 & 19.7 \\
                                 &               & 180   & 21.9 & 22.6 & 21.6 &
20.9 & 19.6 \\
\noalign{\smallskip}
\hline
\end{tabular}
\tablefoot{\\
\tablefoottext{a}{Solar radio flux measured at $10.7$\,cm in sfu.}\\
\tablefoottext{b}{See Table~\ref{tab:modelpar} for model parameters.}
}
\end{table*}

To perform a more quantitative comparison of our sky model and observed data,
and for an easier comparison to other studies, we derived magnitudes from
broad-band filter fluxes. Specifically, we used the standard photometric system
consisting of $U$, $B$, $V$, $R$, and $I$ (see Bessell \cite{BES90}). For the
Patat (\cite{PAT08}) data and their corresponding sky models, only a
grism-dependent subset of filters can be taken. The spectrum needs to cover
$\ga 90$\% of the filter curve. An exception is the $U$ filter, which extends
below $0.365$\,$\mu$m, the lower wavelength limit of the bluest spectral set-up
(see Sect.~\ref{sec:data}). Therefore, the results for this filter are
questionable.

Figure~\ref{fig:magdiff} and Table~\ref{tab:magdiff} show the results of the
computations for the full sample of FORS\,1 spectra and a subsample with the
Moon below the horizon (cf. Sect.~\ref{sec:spec}). Only a small fraction
($\sim 1/4$) of the spectra contribute to the mean values and standard
deviations of the $U$ and $B$ filters. Most spectra cover the range of the $V$,
$R$, and $I$ filters. For both samples and all filters, the mean magnitude
differences are small ($\la 0.05$\,mag). Even though the sky model tends to be
slightly fainter at the blue end and slightly brighter at the red end, there
are no serious systematic deviations from the observed spectra (cf.
Sect.~\ref{sec:spec}). The standard deviations $\sigma_{\Delta m}$ are relatively
similar for the different filters. For the full sample, they range from
0.19\,mag for the $V$-band to 0.24\,mag for the $I$ band. Consequently, the
accuracy of the sky model is in the order of about 20\%. These values can be
compared to the magnitude variations in the measured data $\sigma_{\rm obs}$,
which correspond to $\sigma_{\Delta m}$ for an optimum sky model of
time-invariant flux. For the full sample, the deviations are distinctly larger.
They range from 0.37 to 0.56\,mag. These large deviations are mainly due to the
observations with a strong moonlight contribution (see Sect.~\ref{sec:moon}).
For dark-time conditions, $\langle \Delta m \rangle$ ranges from 0.17\,mag for
$B$ to 0.21\,mag for $U$, and $\sigma_{\Delta m}$ indicates differences between
0.18\,mag for $B$ and 0.29\,mag for $I$. In particular, for the bands $V$ to
$I$, a time-invariant sky model would be significantly worse than our sky model
(see also Sect.~\ref{sec:etc}). This shows the improvements from a variable
zodiacal light, airglow lines, and airglow continuum model.

Finally, we compare the sky model to literature data. Magnitudes in the five
broad-band filters $U$ to $I$ were calculated using standardised observing
conditions. Specifically, zenith, New Moon, ecliptic pole, annual average, and
mean solar activity, i.e. 130\,sfu (see Sect.~\ref{sec:linevar}), were assumed.
The resulting magnitudes are provided in Table~\ref{tab:refmag}. For a better
comparison with published data, the table also contains sky model results for
90 and 180\,sfu. The reference data in Table~\ref{tab:refmag} originate from
Mattila et al. (\cite{MAT96}) for La~Silla, Walker (\cite{WAL87}) and
Krisciunas et al. (\cite{KRI07}) for Cerro Tololo, Benn \& Ellison
(\cite{BEN98}) for La~Palma, and Patat (\cite{PAT03}, \cite{PAT08}) for Cerro
Paranal. In general, there is a good agreement between the sky model and
observed sky brightnesses. By comparing magnitudes with similar average solar
radio fluxes, we find typical deviations in the order of 0.1\,mag. An exception
is the $U$ filter with typical differences of about 0.3\,mag, which could be
explained by the lower quality of the sky model at near-UV wavelengths due to
the lack of data (see Sect.~\ref{sec:data}). Also, it is not clear to what
extent components like zodiacal light or scattered starlight, which
significantly affect short wavelengths, contribute to the observed data. The
corresponding intensities can differ from those of the sky model for the
assumed ideal conditions. The published photometry appears to confirm the lack
of reproducibility of $U$-band magnitudes. Unexpectedly, the sky brightness
seems to increase with decreasing solar activity. It should be noted that
photometric samples are often too small to cover all of the airglow
variability, which causes additional uncertainties. For the airglow continuum,
which is a major contribution to the night-sky brightness in the optical, the
intensity significantly depends on the geographic latitude (see Davis \& Smith
\cite{DAV65}). In summary, we can assume that our sky model calculates reliable
night-sky brightnesses, at least for the levels of solar activity that could be
analysed.

A comparison of the sky model with other studies can also be enlightening in
terms of the night-sky variability, particularly the striking dependence
of the sky brightness on solar activity (Walker \cite{WAL88}; Mattila et al.
\cite{MAT96}; Krisciunas \cite{KRI97}; Patat \cite{PAT03}, \cite{PAT08}). For
the range of photometric fluxes in $UVI$ under dark-time conditions at zenith,
Patat (\cite{PAT08}) found 0.9, 1.3, and 1.7\,mag. For the same filters, he
estimated a flux increase of 0.6, 0.3, and 0.2\,mag for a change in the solar
radio flux from 80 to 240\,sfu. In contrast, Walker (\cite{WAL88}) gives
$\Delta B \approx 0.8$\,mag and $\Delta V \approx 1.0$\,mag for a similar solar
activity change. About $0.5$\,mag are reported by Mattila et al. (\cite{MAT96})
for blue wavelengths and by Krisciunas et al. (\cite{KRI97}) for the $V$ band.
For a typical brightness of the zodiacal light of 1.7 times the intensity at
the ecliptic pole and the same range of solar radio fluxes as used by Patat
(\cite{PAT08}), the sky model gives 0.7, 0.5, 0.6, 0.6, and 0.3\,mag for the
$UBVRI$ filters. These values are consistent with the results from the other
studies. The significant discrepancy of 0.3\,mag to the Patat (\cite{PAT08})
result for the $V$ filter might be explained by the relative small sample of
only 148 images and the general difficulties in separating the different sky
radiation components by means of photometric data. For $B$ to $R$, the solar
activity appears to be responsible for about half the night-sky variability.
Since the zodiacal light variations cause a $1\,\sigma$ uncertainty of about
0.1\,mag, as estimated from the FORS\,1 data set, seasonal, nocturnal, and
unpredictable short-term airglow variations should generate most of the
remaining variability in $V$ and the adjacent bands.

\subsection{Comparison to current ETC sky model}\label{sec:etc}

The main motivation for this study is to achieve a significant improvement over
the current ESO ETC for a more efficient use of the VLT time. As discussed in
Sect.~\ref{sec:introduction}, the current sky emission model for optical
instruments consists of the Moon phase dependent, photometric sky brightness
table of Walker (\cite{WAL87}) and an instrument-specific template spectrum
scaled to the flux in the photometric filter closest to the centre of a given
spectral setting. It does not consider any change in the relative contributions
of the different sky emission components, which can be quite strong as
discussed in the previous sections. Thus, we can expect that the model
continuum level significantly deviates from the true one, even if the
corresponding filter flux is correct. Our new sky model with variable model
components is an important improvement.

For a more quantitative comparison of the current and our new sky model, we
derived the Walker (\cite{WAL87}) magnitudes for the observing conditions of
the Patat (\cite{PAT08}) FORS\,1 data set (see Sect.~\ref{sec:data}). The
resulting $U$ to $I$ magnitudes can be compared to the measured ones to analyse
the quality of the two models (see also Sect.~\ref{sec:photo}).
Table~\ref{tab:magdiff} exhibits the mean deviation of the Walker model
$\langle \Delta m_\mathrm{w} \rangle$ and the standard deviation of the
difference $\sigma_{\Delta m, \mathrm{w}}$ for the full and the dark-time sample.
Since the Walker magnitudes depend on the Moon phase, one can expect that the
model based on his data set would have a lower scatter than an optimum
time-invariant model given by $\sigma_{\rm obs}$. Except for the $U$ band, which
is difficult to compare (see Sect.~\ref{sec:photo}), this is true. For the full
data set, the standard deviations are lower by 0.04 ($I$) to 0.15\,mag ($V$).
However, this improvement is small compared to the corresponding results for
our sky model. For the $B$ and $V$ bands $\sigma_{\Delta m, \mathrm{w}}$ is more
than twice as large as $\sigma_{\Delta m}$. Moreover, there are significant
systematic mean deviations of 0.1 to 0.4\,mag. For our sky model, the maximum
value is 0.05\,mag. Even though this benefits from the use of the same data
for the derivation of the airglow line and continuum model, the offset with the
Walker model is striking. It can be explained by the fact that the Moon phase
is the only variable parameter. Therefore, the systematic changes of other
parameters can cause significant magnitude offsets. For example, the fainter
magnitudes of the Walker model can partly be explained by the very low solar
activity (about 90\,sfu), whereas the Patat (\cite{PAT08}) data set has a mean
solar radio flux of 153\,sfu.

The dark-time data provided by Table~\ref{tab:magdiff} show a decrease of
$\sigma_{\Delta m, \mathrm{w}}$ by 0.02 ($B$) to 0.08\,mag ($R$) compared to the
values for the full sample. Excluding the $U$ band, the typical scatter of the
Walker model is about 0.3\,mag. This is worse than the results for our sky
model and even a time-invariant model. For the $B$ band, the difference is more
than 0.2\,mag. For the $I$ band, the situation is better, since the Walker
model almost reaches the standard deviation of the time-invariant model.
Nevertheless, $\sigma_{\Delta m}$ of our sky model is still about 0.1\,mag lower.
The unsatisfactory performance of the Walker model is caused by the
unconsidered, varying amount of scattered moonlight during a Moon phase. Even
for phases close to Full Moon, dark-time observations are possible. For this
reason, the scatter of the Walker model is larger than for a time-invariant
model. The difference depends on the average contribution of scattered
moonlight to the sky brightness, which is higher at shorter wavelengths. This
effect causes an overestimation of the sky brightness for dark-time
observations, which is indicated by a comparison of $\langle \Delta m_\mathrm{w}
\rangle$ for the full and dark-time samples.

In summary, we can state that our sky model performs better than the current
ESO ETC sky model for the optical by about several tenths of a magnitude. The
quality difference should be even higher for a single wavelength instead of a
broad-band filter. Even if the input parameters of our new sky model (see
Table~\ref{tab:modelpar}) could not optimally be constrained in the planning
phase, due to uncertainties in the scheduling of the observations, there would
be a significant improvement for the prediction of the signal-to-noise ratio
for astronomical observations. This justifies the effort in developing an
advanced sky model.

\section{Summary and outlook}\label{sec:conclusions}

In this paper, we presented a sky model for the ESO observing site at Cerro
Paranal to help predict the effects of the Earth's atmosphere on astronomical
observations in the optical. This sky model includes atmospheric extinction,
scattered moonlight, scattered starlight, zodiacal light, and airglow emission
lines and continuum.
\begin{itemize}
\item[--] The atmospheric extinction is characterised by three components:
Rayleigh scattering, aerosol extinction, and molecular absorption. For
the former, we use the formula given by Liou (\cite{LIO02}). The aerosol
extinction is based on the parametrisation of Patat et al. (\cite{PAT11}) for
Cerro Paranal. The molecular absorption is calculated by the radiative transfer
code LBLRTM for atmospheric profiles optimised for the observing site.
\item[--] Scattered moonlight is derived from the Krisciunas \& Schaefer
(\cite{KRI91}) model for the $V$ band, which is extrapolated for the entire
optical range using the solar spectrum and the Cerro Paranal extinction curve.
\item[--] For scattered starlight, a mean spectrum was derived by 3D single
scattering calculations with multiple scattering corrections, using the Toller
(\cite{TOL81}) integrated starlight distribution and the mean spectrum of
Mattila (\cite{MAT80}).
\item[--] The zodiacal light is computed based on the model given by Leinert
et al. (\cite{LEI98}). Scattering of the zodiacal light in the Earth's
atmosphere is also treated with 3D scattering calculations.
\item[--] Airglow emission lines and continuum are highly variable. We used
a set of 1186 VLT FORS\,1 sky spectra (Patat \cite{PAT08}) taken during a wide
range of conditions to parametrise a model for five line and one continuum
variability classes. The line wavelengths and intensities are based on the
atlas of Hanuschik (\cite{HAN03}), improved by the results of the variability
analysis. For the airglow scattering, 3D calculations were used as well.
\end{itemize}

After summing over all the components, we then compared the sky model with
observed sky spectra. We found that our model is accurate to about 20\%. This
could only be achieved by having variable model components, especially the
airglow model, which is a significant improvement over previous models. The
mean sky brightnesses and variations derived from our model are in good
agreement with results from other photometric studies. Compared to the current
ETC sky model, the accuracy is increased by several tenths of a magnitude.

There are new results from this sky model study:
\begin{itemize}
\item[--] We found scattering scaling relations for the extended radiation
sources zodiacal light and airglow. These are new simple relations to
derive effective extinction curves. The corresponding correction factors for
the point-source extinction curve only depend on the unextinguished
line-of-sight intensity of the radiation source.
\item[--] We characterised the variability of the airglow line and continuum
emission at Cerro Paranal based on several input parameters, such as period of
the year, time of night, and solar radio flux.
\item[--] We derived long-term mean airglow line and continuum intensities.
For the latter, we benefit from the modelling of the intensity of the other sky
model components. We could determine the shape of the optical airglow continuum
with high accuracy. For the controversial wavelength range longwards of
0.7\,$\mu$m, we find a smooth intensity increase. This could be produced by
chemical reactions different from those dominating the shorter wavelengths. The
well-known peak at 0.6\,$\mu$m was identified as strongly varying.
\end{itemize}

The model presented in this paper is aimed at the ETC for optical VLT imagers
and spectrographs. In the near/mid-IR, thermal emission from the telescope,
airglow lines (mainly OH and O$_2$ bands) and continuum and, in particular,
emission/absorption bands from greenhouse gases, originating mainly in the
lower atmosphere, are dominating the sky background of ground-based
observations. The scattering-related components become negligible. The change
in significance between the components as well as the necessity to use IR data
sets for the analysis requires different modelling techniques to achieve
reliable ETC predictions in the IR. The IR sky radiation model will be
described in a forthcoming paper.

The sky model can also be used as a basis for procedures aiming at
reconstructing data by removing the sky signature. This purpose requires the
adaption of the sky model to enable the fitting of observed data. Preliminary
studies show promising results for this kind of application, which will also be
presented in future papers.

\begin{acknowledgements}
We are indebted to F. Patat for providing his FORS\,1 data set and his LBLRTM
set-up for the calculation of the best-fit Paranal extinction curve. We thank
A. Seifahrt for his support and helpful comments on the use of LBLRTM and the
creation of suitable atmospheric profiles. We are grateful to R. Hanuschik for
making his UV spectra available to us. Thanks also go to P. Ballester for his
input to the paper and his prudent management of the project on ESO side and
A. Smette for fruitful discussions. Finally, we thank the anonymous referee for
his detailed and helpful comments. This study was carried out in the framework
of the Austrian ESO In-Kind project funded by BM:wf under contracts
BMWF-10.490/0009-II/10/2009 and BMWF-10.490/0008-II/3/2011. This publication is
also supported by the Austrian Science Fund (FWF).
\end{acknowledgements}

\end{document}